\documentclass[%
 reprint,
superscriptaddress,
showpacs,preprintnumbers,
 amsmath,amssymb,
 prb,
]{revtex4-1}

\usepackage{graphicx}% Include figure files
\usepackage{dcolumn}% Align table columns on decimal point
\usepackage{bm}% bold math
\usepackage{color}
\usepackage{bm}
\usepackage{amsmath}
\newcommand{\LSCO}{La$_{2-x}$Sr$_x$CuO$_4$}
\newcommand{\YBCO}{YBa$_2$Cu$_3$O$_y$}
\newcommand{\Hg}{HgBa$_2$CuO$_{4 + \delta}$}
\newcommand{\Tl}{Tl$_2$Ba$_2$CuO$_6$}
\newcommand{\EuLSCO}{La$_{2-x-y}$Eu$_y$Sr$_x$CuO$_4$}

\newcommand{\ie}{{\it i.e.}}
\newcommand{\eg}{{\it e.g.}}

\newcommand{\Tc}{$T_{\rm c}$}

\newcommand{\Tmin}{$T_{\rm min}$}
\newcommand{\Tmax}{$T_{\rm max}$}
\newcommand{\Tstar}{$T^\star$}

\newcommand{\Hvs}{$H_{\rm vs}$}
\newcommand{\Hc}{$H_{\rm c2}$}
\newcommand{\Hcstar}{$H_{\rm c2}^\star$}
\newcommand{\Hstar}{$H^\star$}

\newcommand{\Nqp}{$N_{\rm qp}$}
\newcommand{\Nsc}{$N_{\rm sc}$}
\newcommand{\NbSi}{Nb$_{\rm x}$Si$_{\rm 1-x}$}

\newcommand{\RH}{$R_{\rm H}$}
\newcommand{\xc}{$x_{\rm c}$}
\newcommand{\pc}{$p_{\rm c}$}
\newcommand{\LB}{$\ell_{\rm B}$}
\newcommand{\vF}{$v_{\rm F}$}
\newcommand{\alphaxy}{$\alpha_{\rm xy}^{\rm sc}$}

\begin{document}

\preprint{APS/123-QED}

%%%%%%%%%%%%%%%%%%%%%%%%%%%% TITLE

\title{Nernst effect in the electron-doped cuprate superconductor Pr$_{2-x}$Ce$_x$CuO$_4$~: 
\\ Superconducting fluctuations, 
upper critical field \Hc, and the origin of the \Tc~dome}

%%%%%%%%%%%%%%%%%%%%%%%%%%%% AUTHORS

\author{F.~F.~Tafti}
\email{Fazel.Fallah.Tafti@USherbrooke.ca}
\affiliation{D\'epartement de physique \& RQMP, Universit\'e de Sherbrooke, Sherbrooke, Qu\'ebec, Canada J1K 2R1}

\author{F.~Lalibert\'e}
\affiliation{D\'epartement de physique \& RQMP, Universit\'e de Sherbrooke, Sherbrooke, Qu\'ebec, Canada J1K 2R1}

\author{M.~Dion}
\affiliation{D\'epartement de physique \& RQMP, Universit\'e de Sherbrooke, Sherbrooke, Qu\'ebec, Canada J1K 2R1}

\author{J.~Gaudet}
\affiliation{D\'epartement de physique \& RQMP, Universit\'e de Sherbrooke, Sherbrooke, Qu\'ebec, Canada J1K 2R1}

\author{P.~Fournier}
\affiliation{D\'epartement de physique \& RQMP, Universit\'e de Sherbrooke, Sherbrooke, Qu\'ebec, Canada J1K 2R1}
\affiliation{Canadian Institute for Advanced Research, Toronto, Ontario, Canada M5G 1Z8}

\author{Louis Taillefer}
\email{Louis.Taillefer@USherbrooke.ca}
\affiliation{D\'epartement de physique \& RQMP, Universit\'e de Sherbrooke, Sherbrooke, Qu\'ebec, Canada J1K 2R1}
\affiliation{Canadian Institute for Advanced Research, Toronto, Ontario, Canada M5G 1Z8}

\date{\today}

%%%%%%%%%%%%%%%%%%%%%%%%%%%% ABSTRACT

\begin{abstract}

The Nernst effect was measured in the electron-doped cuprate superconductor Pr$_{2-x}$Ce$_x$CuO$_4$ (PCCO)
at four concentrations, from underdoped ($x=0.13$) to overdoped ($x=0.17$),
for a wide range of temperatures above the critical temperature \Tc.
A magnetic field $H$ up to 15~T was used to reliably access the normal-state quasiparticle contribution to the Nernst signal, \Nqp,
which is subtracted from the total signal, $N$, to obtain the superconducting contribution, \Nsc.
As a function of $H$, \Nsc~peaks at a field \Hstar~whose temperature dependence obeys 
\Hcstar~$\ln(T/$\Tc), as it does in a conventional superconductor like \NbSi.
The doping dependence of the characteristic field scale \Hcstar~-- shown to be closely related to the 
upper critical field \Hc~-- tracks the dome-like
dependence of \Tc, showing that superconductivity is weakened below the quantum
critical point where the Fermi surface is reconstructed,
presumably by the onset of antiferromagnetic order.
Our data at all dopings are quantitatively consistent with the theory of Gaussian superconducting fluctuations,
eliminating the need to invoke unusual vortex-like excitations above \Tc,
and ruling out phase fluctuations as the mechanism for the fall of \Tc~with underdoping.
We compare the properties of PCCO with those of hole-doped cuprates and conclude that
the domes of \Tc~and \Hc~vs doping in the latter materials are also controlled predominantly by phase competition
rather than phase fluctuations.

\end{abstract}


\maketitle

\section{\label{introduction}Introduction}

Cuprate superconductors have attracted enormous attention because they hold the record for the highest critical temperature
\Tc, which can be as high as 164~K -- halfway to room temperature. \cite{gao_superconductivity_1994}
As a function of doping, \Tc~displays a dome-like dependence, reaching a maximal value at some optimal doping.
A fundamental question is:
Why does \Tc~not continue to rise with underdoping?
A long-held scenario is that the pairing strength (and superconducting gap magnitude) does continue to rise,
but the critical temperature \Tc~for long-range coherence falls because of increasingly
strong fluctuations in the phase of the superconducting order parameter. \cite{Emery_1995}
So the underlying strength of superconductivity would become greater than suggested by the 
maximal (optimal) value of \Tc, and finding ways to increase phase rigidity could further increase
the maximal  \Tc.

The main experimental support for this phase fluctuation scenario came from the observation of
a sizable Nernst signal above \Tc~in underdoped cuprates such as \LSCO~(LSCO). \cite{xu_vortex-like_2000, wang_nernst_2006}
The Nernst effect -- the transverse thermo-electric response to a magnetic field -- 
%is typically
%small in ordinary metals due to Sondheimer cancellation \cite{sondheimer_theory_1948}, but
is large in the vortex-liquid state of type II superconductors, due to the motion of vortices. \cite{vidal_low-frequency_1973}
Consequently, the observation of a large Nernst signal well above \Tc~in hole-doped cuprates was  attributed to short-lived 
vortex excitations above \Tc. \cite{xu_vortex-like_2000}  
In this picture, Cooper pairs with a finite gap in their excitation spectrum survive to temperatures as high as $T \simeq 3$~\Tc.
Defining the upper critical field \Hc~needed to suppress superconductivity as the field where the Nernst signal vanishes,
\Hc~was found to increase with underdoping, and this was taken as evidence of a rising gap.\cite{wang_dependence_2003}
A paradigm was born: 
while the superconducting gap and the upper critical field increase with underdoping, \Tc~decreases due to phase fluctuations.

In recent years, it was shown that the three basic assumptions underlying this interpretation of the Nernst response in cuprates are invalid.
The first assumption was that the quasiparticle contribution to the measured Nernst signal, \Nqp, is negligible,
so that all of the signal can be attributed to superconducting fluctuations.
It has since become clear 
that \Nqp~can in fact be large in a variety of strongly correlated metals,\cite{behnia_nernst_2009} 
including cuprates.\cite{cyr-choiniere_enhancement_2009}
For example, Nernst measurements in \YBCO~(YBCO) and \Hg~(Hg1201) reveal a large negative \Nqp, 
easily disentangled from \Nsc~because its sign is opposite to that of the superconducting signal. 
\cite{rullier-albenque_nernst_2006, daou_broken_2010, chang_nernst_2010, laliberte_fermi-surface_2011, doiron-leyraud_hall_2013}
In these materials, \Nqp~starts to grow below the pseudogap temperature \Tstar, and it becomes comparable in magnitude to \Nsc~at \Tc.
In the electron-doped material PCCO, the two contributions are again readily resolved, even though both are positive in this case,
because \Nqp~exhibits a peak at high temperature while \Nsc~peaks at \Tc, which is relatively low in this material. \cite{balci_nernst_2003, li_normal-state_2007}
A similar two-peak structure is  observed in the hole-doped material 
\EuLSCO~(Eu-LSCO). \cite{cyr-choiniere_enhancement_2009}
In LSCO, the material on which the early studies were based, \Nqp$(T)$ is very similar to that of Eu-LSCO,
but, because \Tc~is higher in LSCO, its peak now merges with the peak in \Nsc$(T)$,
and hence the two contributions are more difficult to disentangle. \cite{cyr-choiniere_enhancement_2009}

The second assumption is that \Nsc$(H)$ vanishes above \Hc.
Nernst measurements on the conventional superconductor \NbSi~have revealed that a superconducting signal 
can persist to fields as high as $H \simeq 4$~\Hc~(and to temperatures as high as $T \simeq 30$~\Tc). \cite{pourret_observation_2006, pourret_length_2007}
This countered the notion that 
superconducting fluctuations
do not exist above \Hc.
The third assumption was that fluctuations which persist up to 2-3~\Tc~cannot be the usual Gaussian fluctuations
of the superconducting order parameter, and hence these were attributed to unusual vortex-like excitations.
In 2009, two groups arrived at a complete theory of Gaussian fluctuations in a dirty 2D superconductor,
extending earlier work \cite{ussishkin_gaussian_2002} to arbitrary 
temperatures and fields. \cite{serbyn_giant_2009, michaeli_fluctuations_2009}
This theory was able to explain in detail and quantitatively the \NbSi~data, proving that standard fluctuations
can indeed persist up to $ T \gg$~\Tc~and $H \gg$~\Hc.

The failure of these three assumptions and the advent of the new theoretical framework 
imposed 
a complete 
re-examination of the Nernst effect in cuprate superconductors.
This started with a study of the hole-doped cuprate Eu-LSCO, in which \Nsc$(T,H)$~was shown to behave 
in the same way as it does in \NbSi~and to agree with Gaussian theory. \cite{chang_decrease_2012}
The characteristic field \Hcstar~extracted directly from the data 
%-- which we show to be equal to \Hc~--
-- which we show here to be approximately equal to \Hc~--
was found to be very low in the underdoped regime.
The authors also re-analyzed the published Nernst data on other hole-doped cuprates to show that \Hcstar~in fact decreases 
with underdoping,\cite{chang_decrease_2012}
 in contrast to the prior report of an increasing \Hc. \cite{wang_dependence_2003}
However, because the Eu-LSCO study was limited to dopings on the underdoped side of the 
\Tc~dome, it did not allow for the ultimate test: to compare the nature of superconducting fluctuations on the two sides of the dome,
and see whether there is a fundamental difference, or not.

In this Article, we report a study of superconducting fluctuations in the electron-doped cuprate PCCO that extends
across the phase diagram,
allowing us to compare both sides of the dome.
This cuprate offers a major advantage in that its critical magnetic 
field at all dopings is low enough that
fluctuations can be fully suppressed by applying only 15 T, thereby allowing a careful extraction of the underlying 
quasiparticle contribution, \Nqp. This is essential for a detailed analysis of \Nsc.
We find that \Nsc~obeys Gaussian fluctuation theory quantitatively -- compelling evidence that 
fluctuations in this cuprate are not unusual, 
on either side of the dome.

We extract \Hcstar~directly from the \Nsc~data and find that it tracks \Tc~as a function of doping,
with \Hcstar~and \Tc~both showing a dome 
that peaks at the same critical doping.
This doping is where the Fermi surface of PCCO is known to undergo a reconstruction,\cite{dagan_evidence_2004} 
due to the onset of a competing phase that breaks translational symmetry,
presumably antiferromagnetic order.\cite{lin_theory_2005}
We conclude that superconductivity in electron-doped cuprates 
weakens below optimal doping not because of superconducting phase fluctuations
but because of phase competition. 
Comparing with the properties of hole-doped cuprates, we argue that the same conclusion applies for hole-doped materials.

\section{\label{previous} Previous work}

The Nernst effect in PCCO has been measured previously,
on thin films with $0.13 \leq x \leq 0.19$.
%in magnetic fields up to 9~T. 
\cite{balci_nernst_2003, li_normal-state_2007,fournier_thermomagnetic_1997}
A sizable \Nsc~was detected above \Tc~in underdoped samples, but {\it not} in overdoped samples.
As a result, Gaussian theory was tentatively ruled out and the
 signal was attributed to phase fluctuations which go away with overdoping. 
In our study, an improved signal-to-noise ratio and a higher 
magnetic field allow us to clearly 
detect \Nsc~in the overdoped regime.
We find that there is in fact no qualitative difference 
between overdoped and underdoped behavior.
For $x \leq 0.15$, our data are consistent with previous data.

\section{\label{experiments}Methods}

We measured four Pr$_{2-x}$Ce$_x$CuO$_{4-\delta}$ thin films 
%with $x=0.13$, 0.14, 0.15, and 0.17 
to cover underdoped ($x=0.13,\, 0.14$), optimally doped ($x=0.15$), 
and overdoped ($x=0.17$) compositions.
The epitaxial thin films with [001] orientation were grown by pulsed laser deposition on LSAT substrates
[(LaAlO$_3$)$_{0.3}$(Sr$_2$AlTaO$_6$)$_{0.7}$]
 using Cu-rich targets to eliminate parasitic phases.\cite{roberge_improving_2009} 

Typical sample dimensions are $3\times 2$ mm, with a thickness of $3000~\textrm{\AA}$.
The Nernst effect was measured using one heater, one differential thermocouple, and one absolute thermocouple.\cite{bougrine_highly_1995}
We used  non-magnetic type E thermocouples made of chromel and constantan wires.
Resistivity measurements were performed in a Quantum Design PPMS.
The electrical and thermal currents were 
%made to flow in
applied along the basal plane of the tetragonal crystal structure,
and the magnetic field was always applied perpendicular to the basal plane, {\it i.e.} along the $c$ axis.
The critical temperature \Tc~is defined as the temperature where the resistivity goes to zero; the values are listed in Table~\ref{scaling}.

%%%%%%%%%%%%%%       TABLE I       %%%%%%%%%%%%%%%%%%%%%%%%%%%%%%%%%%%%

\begin{table}[b]
\caption{\label{scaling} 
Fundamental parameters of superconducting PCCO, as a function of electron concentration $x$: 
critical temperature \Tc, defined as the temperature below which the resistance is zero at $H=0$; 
\Hvs(0), the zero-temperature value of the vortex-solid melting field \Hvs($T$),
defined as the magnetic field below which the resistance is zero;
characteristic magnetic field scale \Hcstar, obtained from the superconducting Nernst signal above \Tc~(see Eq.~\ref{Hstar}).\\}

\begin{tabular}{c c c c}
\hline
$x$&
\Tc&
\Hvs(0)&
\Hcstar
\\
  &
\textrm{(K)}&
\textrm{(T)}&
\textrm{(T)}
\\
\hline
\hline
0.13 & ~8.8~$\pm$~0.3 & ~3.4~$\pm$~0.3 & 2.1~$\pm$~0.2  \\
0.14 & 17.4~$\pm$~0.1 & 10.1~$\pm$~0.5 & 4.3~$\pm$~0.3  \\
0.15 & 19.5~$\pm$~0.1 & ~8.9~$\pm$~0.4 & 5.4~$\pm$~0.3  \\
0.17 & 13.4~$\pm$~0.1 & ~3.0~$\pm$~0.2 & 3.0~$\pm$~0.3 \\
\hline
\end{tabular}
\end{table}

%%%%%%%%%%%%%%       TABLE        %%%%%%%%%%%%%%%%%%%%%%%%%%%%%%%%%

\section{\label{results}Results}
\subsection{\label{resistivity} Resistivity, \Tc~and \Tmin}

%------------------------------------------- FIGURE 1 ---------------------------------------------------------------------
\begin{figure}
\includegraphics[width=3.5in]{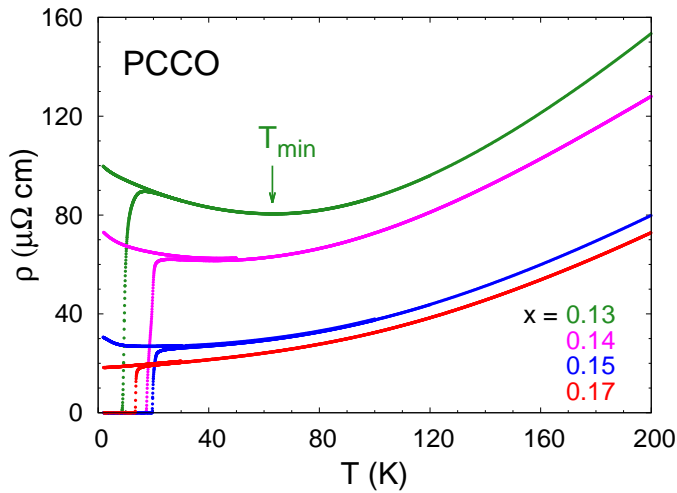}
\caption{\label{RvsT} 
In-plane resistivity of PCCO as a function of temperature for our four thin film samples. 
The data at $H=0$ show the superconducting transition and the data at $H=15$~T show the normal-state behavior. 
For all samples except $x=0.17$, $\rho(T)$ exhibits a minimum, at a temperature \Tmin~(shown here for $x=0.13$; green arrow).
Below \Tmin, $\rho(T)$ rises as $T \to 0$, a signature of Fermi-surface reconstruction.
}
\end{figure}
%-------------------------------------------------------------------------------------------------------------------------

\begin{figure}
\includegraphics[width=3.5in]{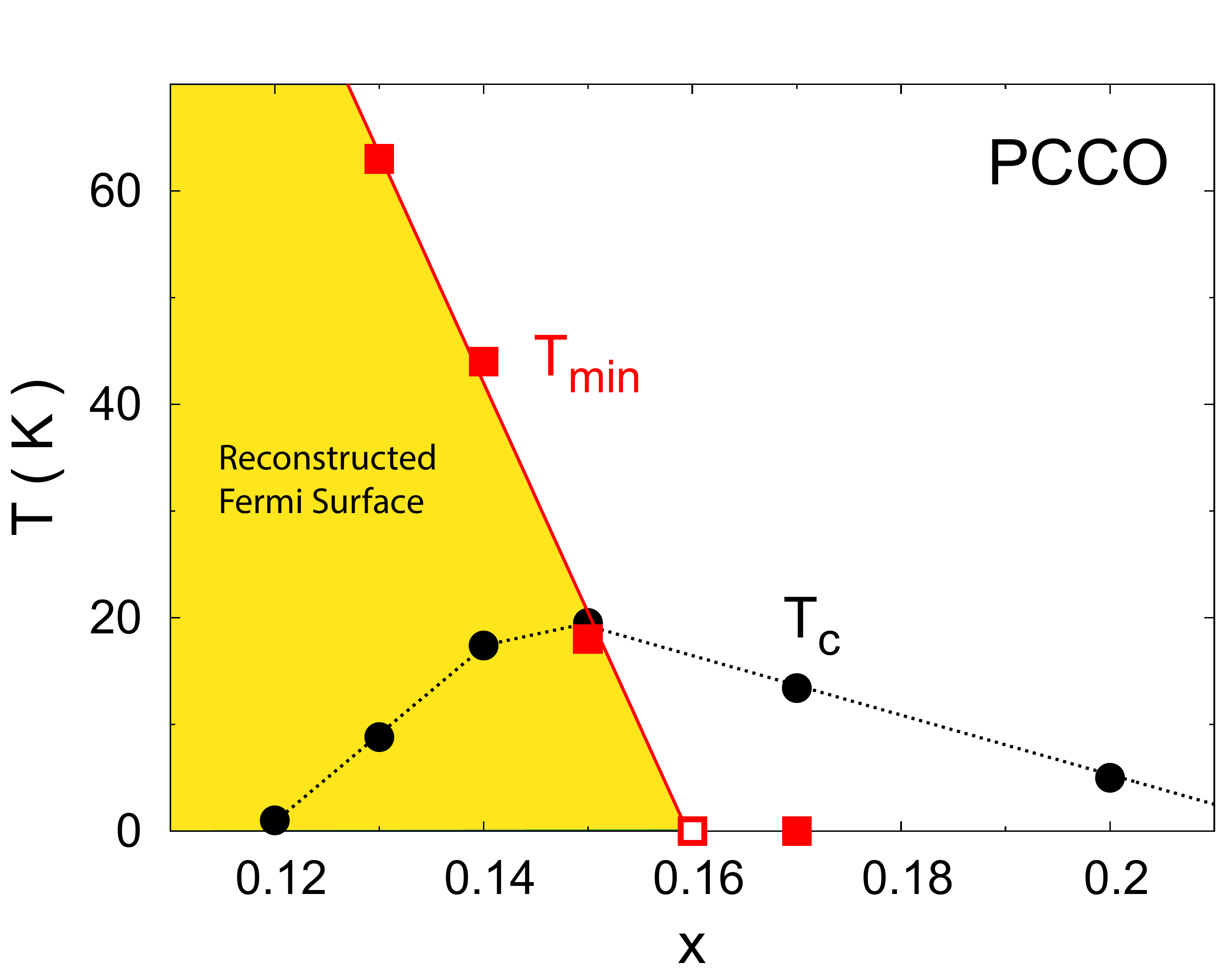}
\caption{\label{TcTminR30} 
Temperature-doping phase diagram of PCCO, showing the superconducting dome delineated by the zero-field
critical temperature \Tc~(black circles).
Also shown is \Tmin~(red squares), the temperature at which the resistivity $\rho(T)$ has a minimum (see Fig.~\ref{RvsT}).
The red line is a linear fit to the \Tmin~data, extrapolated to $T=0$ (open square).
The corresponding doping, \xc~= 0.16, is the quantum critical point below which the 
Fermi-surface reconstruction (FSR) onsets,
in agreement with the critical doping where the normal-state Hall coefficient \RH~at $T \to 0$ exhibits a sharp drop to
negative values. \cite{dagan_evidence_2004}
Throughout our discussion in Sec.~\ref{discussion}, we will use this figure as a map of FSR in PCCO that is derived from resistivity.}

\end{figure}

Figure~\ref{RvsT} displays the resistivity data, $\rho$ vs $T$, for our four samples, at $H=0$ to show the superconducting transition 
and at $H=15$~T to show the 
normal state behavior.
The \Tc~values are plotted in Fig.~\ref{TcTminR30}. 
As in all cuprate superconductors, \Tc~has a characteristic dome-like dependence on doping.
Except at $x = 0.17$, all normal-state resistivity curves $\rho(T)$ show a minimum, at a temperature \Tmin~which 
%shifts up 
increases
with decreasing $x$.
Fig.~\ref{TcTminR30} shows the doping evolution of \Tmin, 
seen to extrapolate to zero at $x = 0.16$. 
This is also the critical doping where the normal-state Fermi surface of PCCO at $T \to 0$ is known to undergo
a reconstruction, detected as a sudden change in the Hall coefficient \RH, going from small and positive 
at $x > 0.16$ to large and negative at $x \leq 0.16$. \cite{dagan_evidence_2004, charpentier_antiferromagnetic_2010}
This Fermi-surface reconstruction (FSR) is attributed to a quantum critical point (QCP) below which some ordered phase sets in, 
at \xc~=~0.16 (in the absence of superconductivity).
Measurements of the Hall coefficient \RH~at $H = 15$~T in our own thin films find that in the low temperature limit, \RH~$> 0$ in our sample with $x = 0.17$ and \RH~$< 0$ in our sample with $x = 0.15$, 
confirming that the normal state QCP at which FSR occurs in our samples is $x_{\rm c} = 0.16 \pm 0.01$, in excellent agreement with prior data. \cite{dagan_evidence_2004, charpentier_antiferromagnetic_2010}
The FSR is also responsible for the upturn in $\rho(T)$ at low $T$.
The resistivity at $x=0.17$ is strictly linear in $T$ below 40 K, in agreement with previous reports.\cite{fournier_insulator-metal_1998, jin_link_2011}
The absence of any upturn shows that $x = 0.17$ is above the critical doping \xc;
the linearity shows that it is close to the QCP.
%
%In Sec.~\ref{FSR}, we elaborate on the QCP and the FSR, 
%phenomena that are fundamental for understanding the 
%superconducting phase diagram.
%
Our resistivity data as presented in Fig.~\ref{RvsT} and \ref{TcTminR30} provide a guide to the reconstruction of the Fermi surface in PCCO and forms the basis of our discussion in Sec.~\ref{FSR}, where we elaborate on the QCP and the FSR, 
phenomena that are fundamental for understanding the superconducting phase diagram.

%------------------------------------------- FIGURE 2 ----------------------------------------------------------------------
%-------------------------------------------------------------------------------------------------------------------------

%------------------------------------------- FIGURE 3 ---------------------------------------------------------------------
\begin{figure}
\includegraphics[width=3.5in]{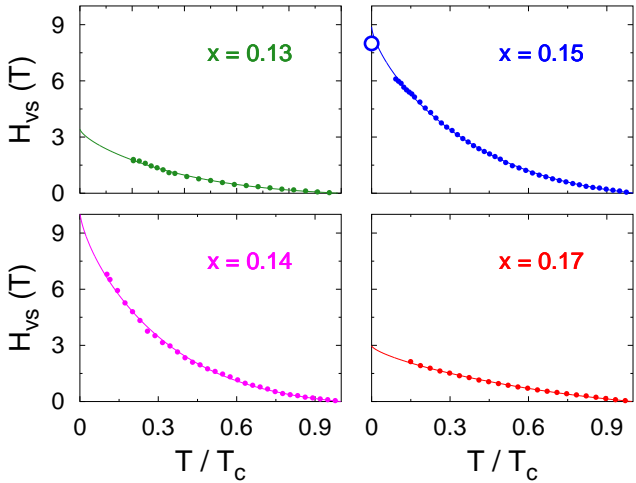}
\caption{\label{T_Hvs} 
Vortex-solid melting lines as a function of temperature, plotted as \Hvs($T$) vs $T$/\Tc, for dopings as indicated. 
\Hvs($T$) is the field below which the sample resistance is zero (full circles).
Solid lines are fits to Eq.~\ref{Hvs}, used to extrapolate \Hvs($T$) to $T=0$ and obtain \Hvs(0), whose value is listed
in Table~\ref{scaling}. 
The open circle in the top right panel marks the value of the upper critical field \Hc~obtained from thermal conductivity data at $x=0.15$ (see Fig.~\ref{Kappa}). 
}
\end{figure}
%-------------------------------------------------------------------------------------------------------------------------

\subsection{\label{vortexmelting} Vortex solid melting field \Hvs}

Having displayed the doping dependence of the critical temperature \Tc,
we now turn our attention to a second fundamental quantity,
the critical magnetic field needed to suppress superconductivity.
We call \Hvs$(T)$~the critical field above which the electrical resistance of the sample
ceases to be zero.
In a type-II superconductor like PCCO, 
this is the field at which the vortex solid melts,
hence the labeling. 
%}
We measured the resistivity of our PCCO films at different temperatures below 
\Tc~to track the temperature dependence of \Hvs~at each doping.
The resulting field-temperature phase diagram is shown in Fig.~\ref{T_Hvs}.
At all dopings, \Hvs~has the typical behavior of cuprate superconductors, 
%with strong upward curvature. 
with positive curvature.

Above the \Hvs$(T)$ line, the electronic state is a vortex liquid. 
The question of where this vortex liquid ends in cuprates has been the subject of much debate.\cite{kivelson_fluctuation_2010}
Recently, it was shown that measurements of the thermal conductivity can be used to answer that question.\cite{grissonnanche_direct_2014}
In three hole-doped cuprates, it was found that there is no vortex-liquid phase at $T \to 0$. \cite{grissonnanche_direct_2014}
In other words, with decreasing field, at $T=0$, vortices appear precisely at \Hvs(0). 
This provides a convenient empirical procedure for determining the upper critical field \Hc, namely \Hc~= \Hvs(0).
Of course, with increasing
temperature, the vortex-liquid phase grows.

%------------------------------------------- FIGURE 4 ---------------------------------------------------------------------
\begin{figure}
\includegraphics[width=3.5in]{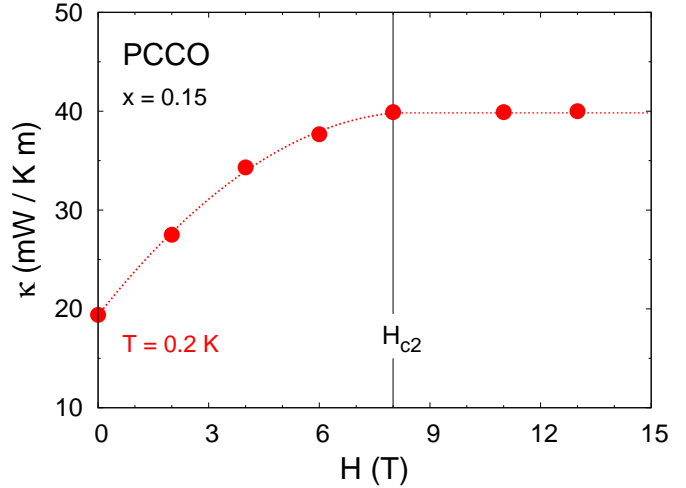}
\caption{\label{Kappa} 
Thermal conductivity $\kappa$ as a function of magnetic field $H$, measured at $T = 0.2$~K 
on a single crystal of PCCO with $x=0.15$ (\Tc~= 20~K) (data from ref.~\onlinecite{hill_breakdown_2001}).
The saturation of $\kappa$ vs $H$ marks the end of the vortex state, providing a direct measurement
of the upper critical field \Hc, defined as the field above which vortices disappear,\cite{grissonnanche_direct_2014}
giving \Hc~$= 8 \pm 1$~T.
}
\end{figure}
%-------------------------------------------------------------------------------------------------------------------------

To extrapolate to $T=0$, we use the standard 
expression for the temperature dependence of \Hvs($T$):\cite{blatter_vortices_1994, houghton_flux_1989, ramshaw_vortex_2012}
\begin{equation}
\frac{\sqrt{b_m(T)}}{1-b_m(T)} \frac{t}{\sqrt{1-t}}\left[\frac{4(\sqrt{2}-1)}{\sqrt{1-b_m(T)}}+1\right] = \frac{2\pi c_L^2}{\sqrt{G_i}}~~~,
\label{Hvs}
\end{equation}
in terms of the reduced field $b_m = $~\Hvs /\Hvs(0)~and reduced temperature $t = T/$\Tc. 
We use the same definitions for the Ginzburg and Lindemann parameters ($G_i$ and $c_L$) as  in ref.~\onlinecite{ramshaw_vortex_2012}.
In Fig.~\ref{T_Hvs}, we see that Eq.~\ref{Hvs} fits the data well and allows us to obtain \Hvs(0),
whose value at each doping is listed in Table I.
The same expression was also found to fit the \Hvs~data of hole-doped cuprates very well, in both
underdoped and overdoped regimes. \cite{grissonnanche_direct_2014, ramshaw_vortex_2012}

We can use existing thermal conductivity data to confirm that \Hc~is indeed equal to \Hvs($T\to0$) in PCCO. 
In Fig.~\ref{Kappa}, we reproduce published data taken at $T~=~0.2$~K on a single crystal of PCCO 
at optimal doping ($x = 0.15$, \Tc~= 20~K). \cite{hill_breakdown_2001}
We see from those data that \Hc~=~8~$\pm~1$~T, in excellent agreement with the value of \Hvs(0)~=~8.9~$\pm~0.4$~T
we obtain by extrapolating \Hvs($T$) to $T=0$ (Fig.~\ref{T_Hvs}).
A similar value for \Hvs(0) is reported in prior studies. \cite{fournier_doping_2003}

%------------------------------------------- FIGURE 5 ----------------------------------------------------------------------
\begin{figure}
\includegraphics[width=3.5in]{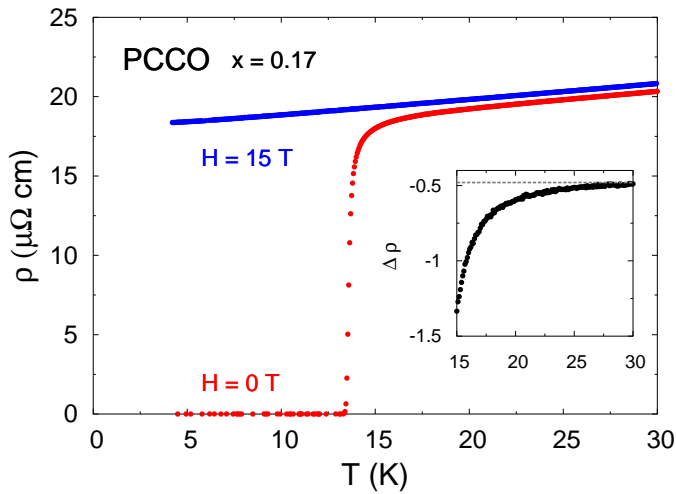}
\caption{\label{rho_0p17} 
Resistivity of our PCCO sample with $x = 0.17$.
The critical temperature is defined as the temperature where the zero-field data (red) goes to zero:
\Tc~$=13.4$~K.
The normal-state behavior is given by the data in a field $H=15$~T (blue), which includes a slight rigid upward shift 
due to positive magnetoresistance.
A regime of paraconductivity is detectable below $T \simeq 2$~\Tc, due to superconducting fluctuations that gradually decrease
$\rho(T)$, before its rapid drop to zero at \Tc.
{\it Inset}:
Zoom on paraconductivity: 
$\Delta \rho \equiv \rho(0) - \rho(15~{\rm T})$ vs $T$.
}
\end{figure}
%-------------------------------------------------------------------------------------------------------------------------

\subsection{\label{nernst} Nernst effect : overdoped sample ($x = 0.17$)}

In presenting our Nernst data on PCCO, we begin with the overdoped sample at $x=0.17$.
There are two reasons for this initial focus.
First, this sample provides a fundamental reference point for all Nernst studies of cuprate superconductors, missing until now.
In an overdoped sample, the Fermi surface is neither reconstructed nor altered by a pseudogap:
it is simply a single large hole-like cylinder, whose area is precisely given by the doping.
In hole-doped cuprates, this $k$-space area is proportional to $1 + p$; \cite{mackenzie_normal-state_1996}
in PCCO, it is proportional to $1 - x$.\cite{lin_theory_2005}
Moreover, in PCCO, we have a particularly simple crystal structure, which is tetragonal
and free of bilayers, buckling, oxygen order or chains. \cite{armitage_progress_2010}
The Nernst data we report here can therefore be regarded as the archetype of an overdoped cuprate,
the property of a single pristine CuO$_2$ plane.
Second, having data in the overdoped regime will enable us to make the first direct comparison
of  superconducting fluctuations on the left and right sides of the \Tc~dome, and establish
the differences, if any.

%------------------------------------------- FIGURE 6 ----------------------------------------------------------------------
\begin{figure}
\includegraphics[width=3.5in]{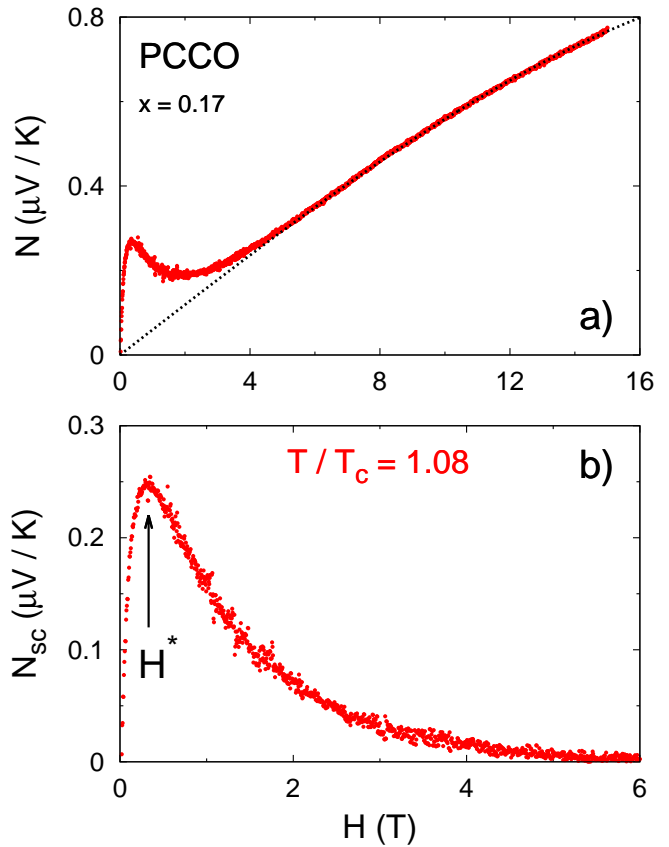}
\caption{\label{NvsH_17} 
Nernst response of PCCO as a function of magnetic field $H$ in our sample with $x = 0.17$, at $T = 1.08$~\Tc.
a)~Raw Nernst signal $N$ vs $H$ (red). The black dotted line is a polynomial fit to the data above 10~T,
of the form \Nqp~$= a(T) H + b(T) H^3$.
\Nqp~is the quasiparticle background of the underlying normal state.
b) Superconducting contribution to the Nernst signal, defined as \Nsc~$\equiv N -$\Nqp~(Eq.~\ref{Ntotal}).
For any given temperature above \Tc,
\Nsc~vs $H$  
exhibits a peak, at a field labelled \Hstar (arrow).
}
\end{figure}
%-------------------------------------------------------------------------------------------------------------------------

In Fig.~\ref{rho_0p17}, we show the in-plane resistivity $\rho(T)$ of our $x=0.17$ sample, below 30~K.
With increasing temperature, $\rho$ rises suddenly at \Tc~$= 13.4$~K. 
Above \Tc, there is a regime of paraconductivity, where superconducting fluctuations
reduce the resistivity from its normal-state value. 
The data at $H=15$~T provide the normal-state reference, {\it modulo} a small rigid shift 
in $\rho(T)$ due to a positive orbital magneto-resistance.
Note that paraconductivity can be seen up to 2 \Tc~or so (see inset of Fig.~\ref{rho_0p17}).

%------------------------------------------- FIGURE 7 ----------------------------------------------------------------------
\begin{figure}
\includegraphics[width=3.5in]{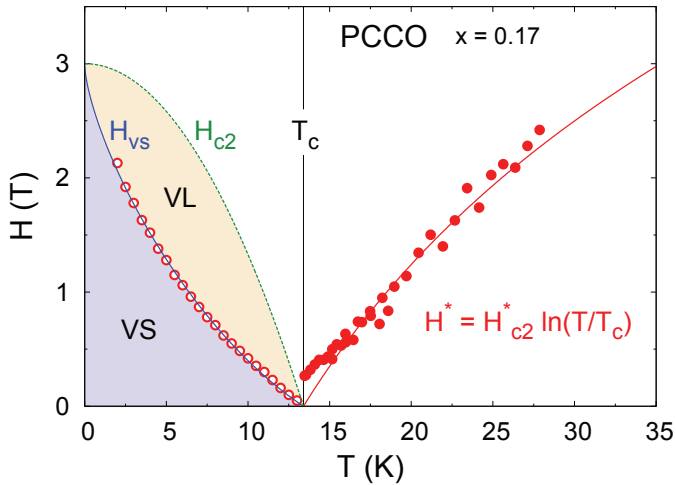}
\caption{\label{Hstar_17_All} 
Magnetic field-temperature phase diagram of PCCO at $x = 0.17$.
The peak field \Hstar~(full red dots) is plotted above \Tc~(vertical black line).
The solid red line is a fit of the \Hstar~vs $T$ data, to the formula 
\Hstar = \Hcstar~$\ln (T/T_{\rm c})$~(Eq.~\ref{Hstar}).
The vortex-solid melting field \Hvs($T$)~(open circles) is the boundary between the vortex-solid phase (VS),
where the electrical resistance is zero, and the vortex-liquid phase (VL), where the resistance is non-zero.
The upper critical field \Hc~(green dotted line) is the boundary between the vortex liquid and the normal state, 
where there are no vortices.
At $T=0$, there is no vortex liquid, since \Hc(0) = \Hvs(0), as reported for
hole-doped cuprates,\cite{grissonnanche_direct_2014}
and shown here for PCCO at $x = 0.15$ (Figs.~\ref{T_Hvs} and~\ref{Kappa}).
The green dotted line is a schematic representation.
Note that the two characteristic field scales for superconductivity, obtained respectively
at $T \to 0$ and at $T >$~\Tc, are equal for $x = 0.17$, namely  \Hc(0) = \Hcstar~$= 3.0$~T (Table~\ref{scaling}).
}
\end{figure}
%-------------------------------------------------------------------------------------------------------------------------

The raw Nernst signal $N$ as a function of field is shown in Fig.~\ref{NvsH_17}a, at $T = 1.08$~\Tc.
$N(H)$ shows an initial rise with a subsequent fall on top of a smoothly increasing background.
The peak at low field is due to superconducting fluctuations, \Nsc, while the background 
is the normal-state quasiparticle signal, \Nqp.
The total Nernst signal is the sum of these two components: 
\begin{equation}
N = N_{\rm sc} + N_{\rm qp}~~~.
\label{Ntotal}
\end{equation}
To establish the background for each isotherm, we fit the data above 10~T to a power law: 
\begin{equation}
N_{\rm qp} = a(T) H + b(T) H^3~~~.
\label{Nbksubt}
\end{equation}
The dotted line in Fig.~\ref{NvsH_17}a is a fit to Eq.~\ref{Nbksubt}. 
We see that it describes the raw data very well from $\sim 6$~T all the way to 15~T.
Note that a cubic term is essential to capture the correct $H$ dependence of \Nqp.
In previous work, limited to lower fields,\cite{li_normal-state_2007} 
\Nqp~was assumed to have a purely linear dependence, but in principle all odd powers of $H$ are allowed by symmetry.

%------------------------------------------- FIGURE 8 ---------------------------------------------------------------------
\begin{figure}
\includegraphics[width=3.5in]{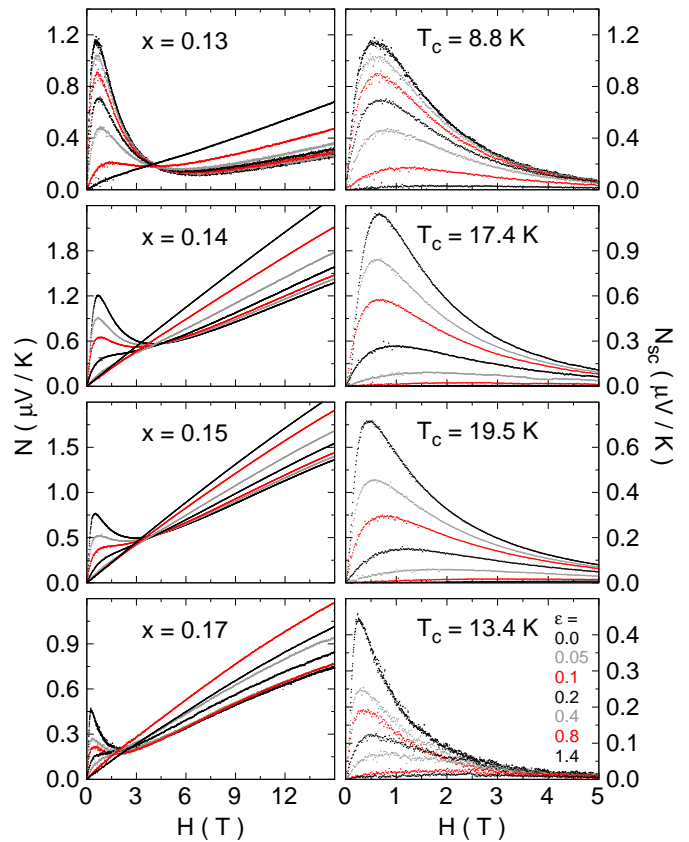}
\caption{\label{NvsH_alldopings}
{\it Left panels}:
Raw Nernst data as a function of field for our 
four samples, at dopings as indicated. 
For each doping, we present seven isotherms, at $\epsilon= 0.0$, 0.05, 0.1, 0.2, 0.4, 0.8, and 1.4,
where $\epsilon \equiv (T-$\Tc)/\Tc, with \Tc~as indicated. 
{\it Right panels}:
Superconducting contribution to the Nernst signal, \Nsc, obtained by subtracting the normal-state background, \Nqp,
as shown in 
Figs.~\ref{NvsH_17}~and~\ref{NvsH_13_14}, according to Eqs.~\ref{Ntotal} and~\ref{Nbksubt}.
Note that with increasing temperature the magnitude of \Nsc~decreases and the peak field \Hstar~moves up.
\Hstar~is plotted vs $\epsilon$ in Fig.~\ref{T_Hstar}.}
\end{figure}
%-------------------------------------------------------------------------------------------------------------------------

%------------------------------------------- FIGURE 9 ----------------------------------------------------------------------
\begin{figure}
\includegraphics[width=3.5in]{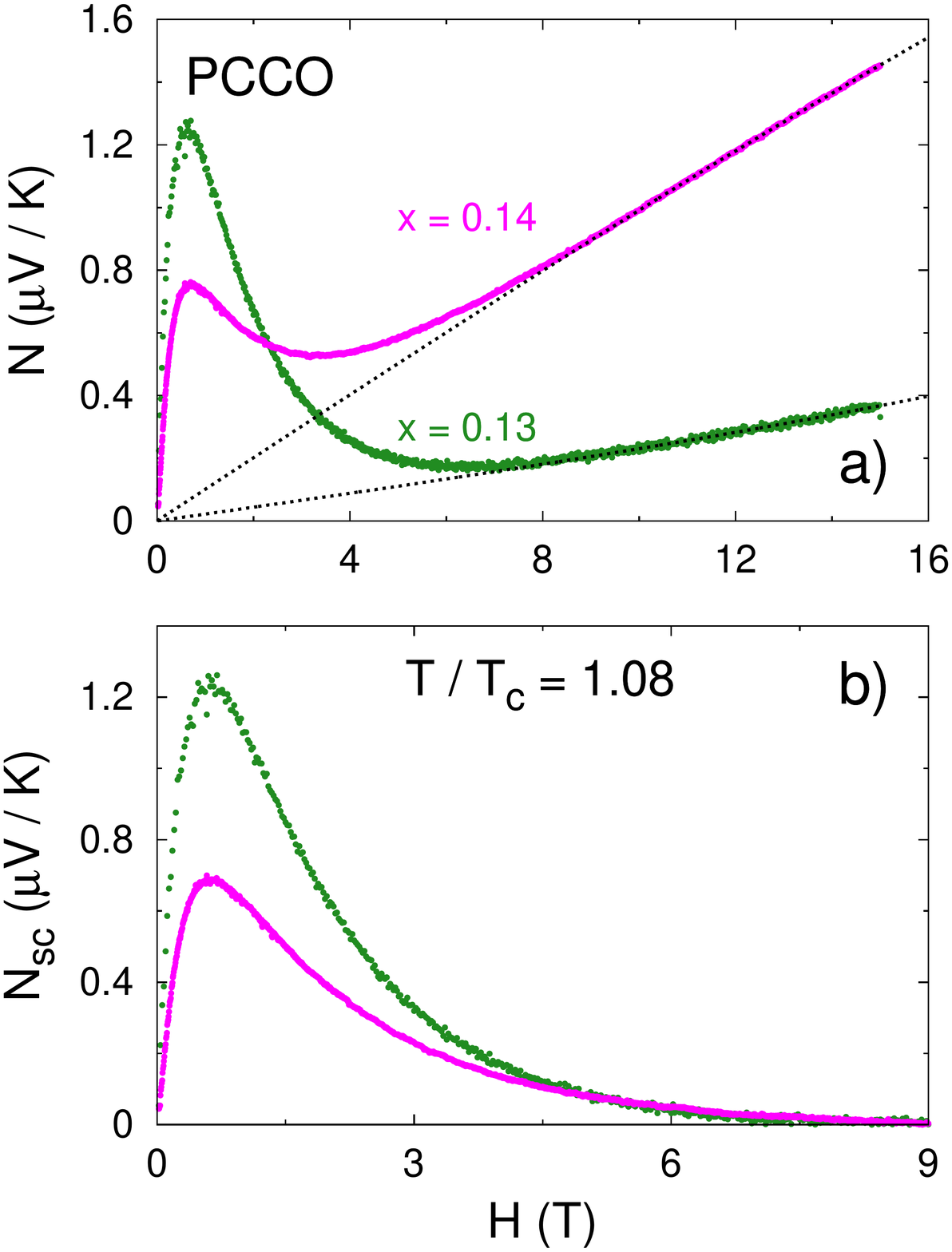}
\caption{\label{NvsH_13_14} 
Nernst response of PCCO vs magnetic field, for $x~=~0.13$ (green) and
$x = 0.14$ (magenta), at $T = 1.08$~\Tc.
a) Raw Nernst signal $N$ vs $H$. The black dotted line is a polynomial fit to the data above 10~T,
of the form \Nqp~$=~a(T) H + b(T) H^3$ (Eq.~\ref{Nbksubt}).
b) Superconducting contribution to the Nernst signal, defined as \Nsc~$\equiv N -$\Nqp~(Eq.~\ref{Ntotal}).
}
\end{figure}
%-------------------------------------------------------------------------------------------------------------------------

In Fig.~\ref{NvsH_17}b, we plot the superconducting signal \Nsc~=~$N -$ \Nqp.
It rises rapidly from zero, goes through a peak, and then decreases gradually, to eventually become vanishingly small at high field, for $H >$~2 \Hc~or so.
(It has been suggested that there may be an intrinsic limit to how high superconducting fluctuations can extend above \Hc~(or \Tc), associated with the uncertainty principle.
\cite{vidal_consequences_2002, soto_experimental_2004})
%In the overdoped $x=0.17$ sample, \Nsc~vanishes at roughly 2\Hc~consistent with theoretical predictions based on uncertainty principle in superconducting fluctuations \cite{vidal_consequences_2002, soto_experimental_2004}}.
This is a  typical signal for a superconductor.
In fact, \Nsc~displays these same features in all superconductors, at $T >$~\Tc.
In particular, there is always a peak field, which we label \Hstar.
We see that \Hstar~is a characteristic field scale for superconductivity in a given material
that is directly and immediately obtainable from the data, with no assumptions and no model or theory.

As a function of temperature, \Nsc~decreases in magnitude, but the peak field  \Hstar~increases.
In Fig.~\ref{Hstar_17_All}, we show on a field-temperature phase diagram the temperature dependence of \Hstar~above \Tc, 
and also \Hvs($T$) measured on the same sample, below \Tc.
In addition, we sketch the temperature dependence of the upper critical field \Hc($T$).
\Hstar($T$) and \Hc($T$) are 
images of each other on either side of \Tc, 
and for this reason  \Hstar~has been called the ``ghost critical field". \cite{kapitulnik_inhomogeneity_1985}
The \Hstar~data can be fit to a logarithmic dependence,
such that 
\begin{equation}
H^\star = H_{\rm {c2}}^\star \ln (T/T_{\rm c})~~~.
\label{Hstar}
\end{equation}
The prefactor \Hcstar~is a single empirical parameter that 
characterizes the strength of superconductivity.

The $ \ln (T/T_{\rm c})$ dependence in Eq.~\ref{Hstar}~was explained intuitively by Pourret \emph{et al.} in the context of \NbSi~thin films.\cite{pourret_length_2007}
They 
proposed
that the crossover from increasing to decreasing \Nsc~occurs because the length scale for superconducting fluctuations at low $H$
is set by the coherence length $\xi(T)$, while at high $H$~it is set by the magnetic length \LB~$=\sqrt{\hbar/eH}$.
\Hstar~would be the field where the two length scales become comparable, \ie~$\xi$(\Hstar)~$\simeq$~\LB(\Hstar).
Since $\xi(T) \propto 1 / \sqrt{\ln (T/T_{\rm c})}$ and \LB(\Hstar)~$\propto 1 / \sqrt{H^\star}$, this yields \Hstar~$\propto \ln (T/T_{\rm c})$.

The one-parameter fit to the $x=0.17$ data in Fig.~\ref{Hstar_17_All} using Eq.~\ref{Hstar} yields \Hcstar~$= 3.0 \pm 0.3$~T (Table~\ref{scaling}).
(Note that the data deviate from the fit close to \Tc~-- an intrinsic effect explained in Sec.~\ref{gaussiancomparison}.)
This value can be compared with our estimate of \Hc~for that same sample, obtained as the $T=0$ limit of the resistive critical field \Hvs($T$),
whose value is \Hvs(0)~$= 3.0 \pm 0.2$~T (see Table~\ref{scaling} and Figs.~\ref{T_Hvs}, \ref{Hstar_17_All}).
We arrive at a useful empirical result: 
the characteristic field scale encoded in superconducting fluctuations above \Tc, when defined as in Eq.~\ref{Hstar},
is equal to the field needed to kill superconductivity at $T=0$.
In other words, we now have a straightforward empirical procedure for measuring the fundamental field 
scale for superconductivity, \Hc, from superconducting fluctuations above \Tc.
Note that this is for a single-band $d$-wave superconductor.

In summary, superconducting fluctuations in overdoped PCCO, at $x=0.17$, are detectable in \Nsc~up to
$T \simeq 2$~\Tc~and $H \simeq 2$~\Hc, and they can be used to measure \Hc.
What is the nature of these fluctuations?
As we show in Sec.~\ref{discussion}, not only does \Nsc~have precisely the field dependence predicted by Gaussian theory,
 its magnitude is in excellent agreement with theoretical expectation. 
We conclude that the superconducting fluctuations of an overdoped cuprate are now well understood.
%}

%------------------------------------------- FIGURE 10 ---------------------------------------------------------------------
\begin{figure}
\includegraphics[width=3.5in]{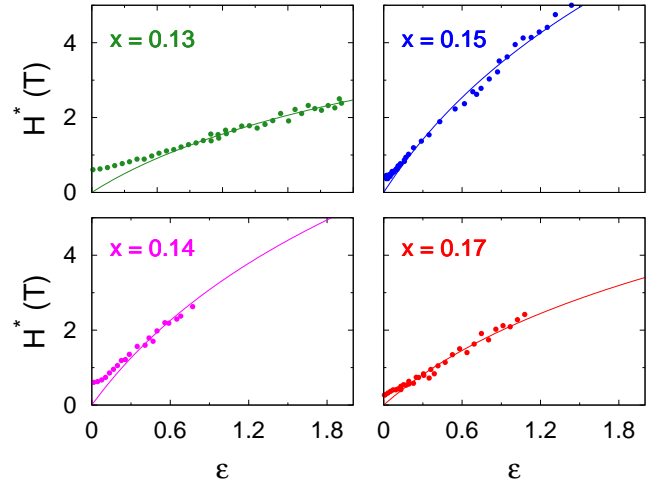}
\caption{\label{T_Hstar} 
Temperature dependence of the peak field \Hstar~in PCCO, at dopings as indicated,
plotted as a function of $\epsilon\equiv(T-$\Tc)/\Tc.
\Hstar~is the field at which \Nsc~peaks, in isotherms of \Nsc~vs $H$ 
as shown in the right panels of Fig.~\ref{NvsH_alldopings}.
The lines are a fit of the data to the function \Hstar = \Hcstar~$\ln (T/T_{\rm c})$~(Eq.~\ref{Hstar}).
The fit allows us to extract a single characteristic field, \Hcstar, from the superconducting fluctuations at each doping,
directly from the data.
The values of the single fit parameter, \Hcstar, are listed in Table~\ref{scaling} and plotted vs $x$ in Fig.~\ref{x_Hnu_Hstar_Hvs}.
At low $\epsilon$, the data deviate from the fit for 
reasons given in Sec.~\ref{gaussiancomparison}.
}
\end{figure}
%-------------------------------------------------------------------------------------------------------------------------

%------------------------------------------- FIGURE 11 -------------------------------------------------------------------
\begin{figure}
\includegraphics[width=3.5in]{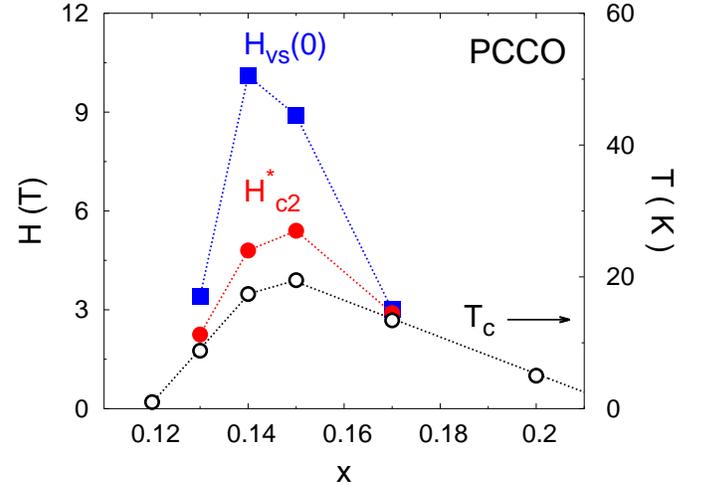}% Here is how to import EPS art
\caption{\label{x_Hnu_Hstar_Hvs} 
The two magnetic field scales of superconductivity in PCCO, plotted as a function of doping~(Table~\ref{scaling}).
\Hvs(0)~(blue squares), the zero-temperature value of the upper critical field,
is obtained by extrapolating to $T=0$ the vortex-solid melting field \Hvs($T$) below which the resistance is zero (see Fig.~\ref{T_Hvs}).
The field scale \Hcstar~(red circles) is obtained from the superconducting fluctuations above \Tc~(see Fig.~\ref{T_Hstar}). 
Note that \Hcstar~=~\Hvs(0)~at $x = 0.17$.
}
\end{figure}
%-------------------------------------------------------------------------------------------------------------------------

%------------------------------------------- FIGURE 12  ---------------------------------------------------------------------
\begin{figure}
\includegraphics[width=3.5in]{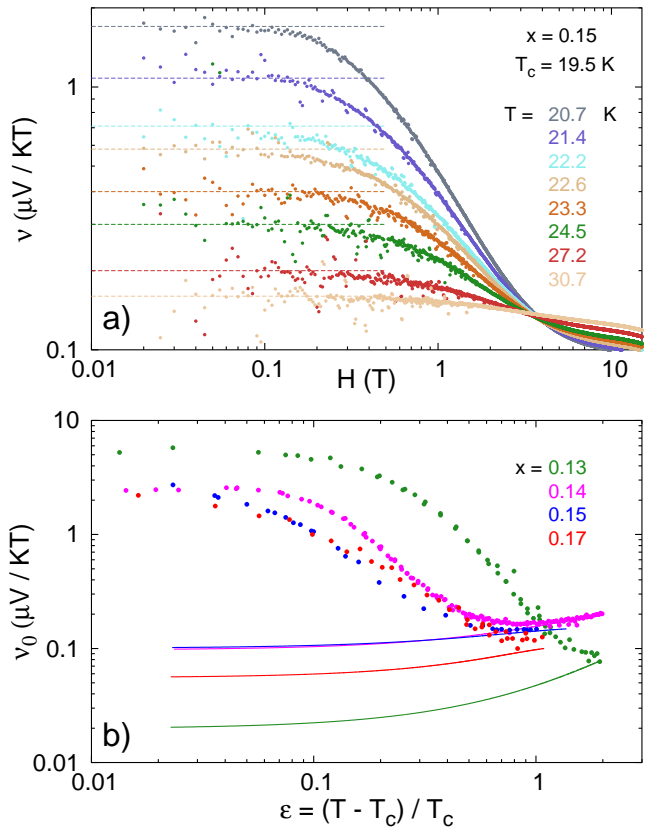}
\caption{\label{nu0} 
a) Raw Nernst coefficient $\nu \equiv N/H$ of PCCO at $x=0.15$, for selected isotherms above \Tc, as indicated. 
In the limit  $H \to 0$, $\nu(H)$ becomes flat, reaching a constant value, $\nu_0$, given by the dashed line.
b) Temperature dependence of $\nu_0$ (dots), at dopings as indicated,
plotted as a function of $\epsilon$.
The solid lines are the normal-state contribution to $\nu_0$, \ie~ $\nu_{\rm qp} \equiv$~\Nqp$/H$ in the limit $H \to 0$.
Note that $\nu_0$~saturates below $\epsilon \simeq 0.1$, as expected from Gaussian theory in the limit  $T \to$~\Tc~(see Sec.~\ref{gaussiancomparison}).
}
\end{figure}
%-------------------------------------------------------------------------------------------------------------------------

%%%%%%%%%%%%%%       TABLE   II     %%%%%%%%%%%%%%%%%%%%%%%%%%%%%%
\begin{table}[b]
\caption{\label{alphatable} 
Superconducting contribution to the off-diagonal Peltier coefficient $\alpha_{\rm xy}$ of PCCO at $H \to 0$, 
given per CuO$_2$ plane by $\alpha_{\rm xy}^{\rm sc} / H = \nu_{0}^{\rm sc} / \rho_{\square}$, 
as a function of $x$.
$\nu_{0}$~is the value of the raw Nernst coefficient $\nu \equiv N / H$ at $H \to 0$ (Fig.~\ref{nu0});
%$\nu_{0}^{\rm qp}$ is the quasiparticle contribution (Fig.~\ref{nu0});
%$\nu_{0}^{\rm sc} = \nu_{0} - \nu_{0}^{\rm qp}$ is the superconducting contribution;
$\nu_{0}^{\rm sc} = \nu_{0} - \nu_{0}^{\rm qp}$ is the superconducting contribution obtained 
by subtracting the quasiparticle contribution $\nu_{0}^{\rm qp}$ (Fig.~\ref{nu0});
$\rho$~is the electrical resistivity at $H=0$ (Fig.~\ref{RvsT}), and 
$\rho_{\square} = \rho / s$, with $s=6.1$~\AA.
All values are measured at $\epsilon = 0.1$~(\ie~at $T = 1.1$~\Tc).\\}
\begin{tabular}{c c c c c}
\hline
$x$ & $\nu_0$ & $\nu_0^{\rm sc}$ & $\rho$ & $\alpha_{xy}^{\rm sc}/H$\\
 & (nV/KT) & (nV/KT) & ($\mu \Omega$ cm)  & (nA/KT) \\
\hline\hline
0.13 & 4560 & 4540   & 65$\>\pm\>$15    & 4.3$\>\pm\>$1.0 \\
0.14 & 1890 & 1790   & 45$\>\pm\>$10  & 2.4$\>\pm\>$0.5 \\
0.15 & 1060 & 960     & 25$\>\pm\>$3  & 2.3$\>\pm\>$0.3 \\
0.17 & 1000 & 940     & 18$\>\pm\>$5  & 3.2$\>\pm\>$0.9 \\
\hline
\end{tabular}

\end{table}
%%%%%%%%%%%%%%%%%%%%%%%%%%%%%%%%%%   TABLE   %%%%%%%%%%%%%%%%%%%%%%%%%%%%%%%%%%%%%

\subsection{\label{nernst} Nernst effect : all dopings ($0.13 \leq x \leq 0.17$)}

The left panels of Fig. \ref{NvsH_alldopings} present a selection of raw Nernst isotherms,
labelled by their reduced temperature $\epsilon \equiv (T - T_{\rm c})/T_{\rm c}$, for each of the four samples.
The superconducting Nernst signal \Nsc~is shown in the corresponding right panels.
Examples of background subtraction are given in Fig.~\ref{NvsH_13_14}, for $x = 0.13$ and $x=0.14$.
As in the $x=0.17$ sample, there is a large positive \Nqp, mostly linear in $H$, but with a small additional $H^3$ term.
The peak field \Hstar~obtained from \Nsc~vs $H$ is plotted vs $\epsilon$ in Fig.~\ref{T_Hstar} for the four dopings.
A fit to Eq.~\ref{Hstar} yields the \Hcstar~values listed in Table~\ref{scaling}, and plotted vs $x$ in Fig.~\ref{x_Hnu_Hstar_Hvs}.
For all $x$, the data deviate from the fit as $T \to$~\Tc, for reasons given in sec.~\ref{gaussiancomparison}.
In Fig.~\ref{nu0}a, we plot the Nernst coefficient $\nu$, defined as $\nu \equiv N / H$, vs $H$.
We see that it is flat at low $H$, {\it i.e.} 
$N$ is linear as $H \to 0$.
In Fig.~\ref{nu0}b, we plot the initial value of $\nu(H)$, which we call $\nu_0$, vs $\epsilon$. 
In Table~\ref{alphatable}, we list the values of $\nu_0$ for the 
four dopings, at $\epsilon = 0.1$, and compare
these to theoretical expectation in Sec.~\ref{gaussiancomparison}.
%}

%%%%%%%%%%%%%%%%%%%%%%%%%%%%%%%%%%%%%%%%%%%%%%%%%%%%%%%%%%%%%%%%%%%%%%
%%%%%%%%%%%%%%%%%%%%%%%%%%%%%%%%%%%%%%%%%%%%%%%%%%%%%%%%%%%%%%%%%%%%%%%
%%%%%%%%%%%%%%%%%%%%%%%%%%%%%%%%%%%%%%%%%%%%%%%%%%%%%%%%%%%%%%%%%%%%%%
%%%%%%%%%%%%%%%%%%%%%%%%%%%            DISCUSSION       %%%%%%%%%%%%%%%%%%%%%%%%%%%%%%%%
%%%%%%%%%%%%%%%%%%%%%%%%%%%%%%%%%%%%%%%%%%%%%%%%%%%%%%%%%%%%%%%%%%%%%%%
%%%%%%%%%%%%%%%%%%%%%%%%%%%%%%%%%%%%%%%%%%%%%%%%%%%%%%%%%%%%%%%%%%%%%%
%%%%%%%%%%%%%%%%%%%%%%%%%%%%%%%%%%%%%%%%%%%%%%%%%%%%%%%%%%%%%%%%%%%%%%%

\section{\label{discussion}Discussion}
Having presented our data for the Nernst signal in PCCO as a function of $H$, $T$ and $x$,
we now examine what they tell us about the nature of the superconducting fluctuations
and the mechanisms that control the strength of superconductivity in cuprates.

%------------------------------------------- FIGURE 13 ---------------------------------------------------------------------
\begin{figure}
\includegraphics[width=3.5in]{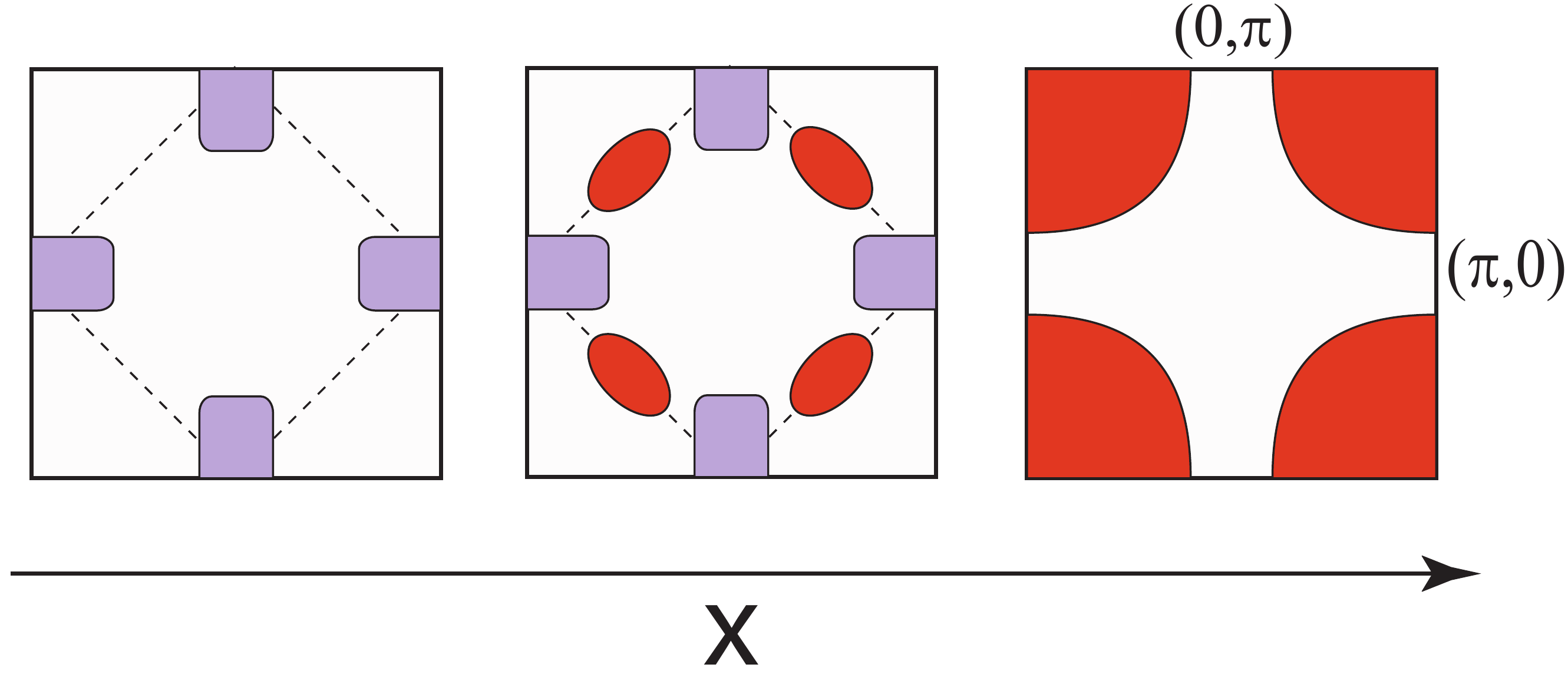}% Here is how to import EPS art
\caption{\label{PCCO_FSR} 
Sketch of the doping evolution of the Fermi surface in PCCO,
based on ARPES measurements performed on NCCO, a closely-related material. \cite{armitage_doping_2002, matsui_evolution_2007}
At high $x$, in the overdoped regime, the Fermi surface is a single large hole-like nearly circular 
Fermi cylinder (drawn in red).
Below a critical doping \xc~$\simeq 0.16$, the Fermi surface undergoes a reconstruction,
into small hole (red) and electron (blue) pockets.
At low $x$, the small hole pockets eventually disappear, leaving only the electron pockets 
centered at $(\pm~\pi, 0)$ and $(0, \pm~\pi)$.
}
\end{figure}
%-------------------------------------------------------------------------------------------------------------------------

\subsection{Fermi-surface reconstruction}
\label{FSR}
To make sense of the doping dependence of superconductivity in PCCO, it is essential
to first describe the underlying normal state and how it evolves with doping.
The key organizing principle is a quantum critical point \xc~at which the Fermi surface undergoes
a major transformation.
The evolution is sketched in Fig.~\ref{PCCO_FSR}.
Above \xc, the Fermi surface of PCCO is a single large closed hole-like cylinder, with a $k$-space area 
given by $1-x$. 
This is confirmed experimentally in several ways. 
First, in the limit of $T=0$ the Hall coefficient \RH~$= + 1 / n e$, where the carrier density $n = 1 - x$ carriers per Cu atom. \cite{dagan_evidence_2004}
Second, the frequency $F$ of quantum oscillations detected in overdoped Nd$_{2-x}$Ce$_x$CuO$_4$~(NCCO), 
a closely related material,
is such that $F = n \Phi_0$. \cite{helm_evolution_2009}
Third, measurements of angle-resolved photoemission spectroscopy (ARPES) in NCCO
see a large closed Fermi surface of the right area. \cite{armitage_doping_2002, matsui_evolution_2007}

In the normal state at $T \to 0$, achieved by applying a field $H >$~\Hc, \RH~undergoes a sudden and dramatic change
below \xc~= 0.16.
It goes from small and positive to large and negative. \cite{dagan_evidence_2004}
ARPES measurements on NCCO reveal a transformation as sketched in Fig.~\ref{PCCO_FSR},
whereby the large hole-like cylinder of the overdoped regime is reconstructed into
two small pockets, respectively located at $(\pi,0)$ and $(\pi/2,\pi/2)$, as the doping is reduced below $x \simeq 0.16$. \cite{matsui_evolution_2007}
This reconstruction is consistent with the observation of 
low-frequency quantum oscillations in NCCO, \cite{helm_evolution_2009}
which reveal the existence of a small closed pocket in the Fermi surface, 
tentatively attributed to the pocket seen by ARPES 
near $(\pi/2,\pi/2)$. 

The evidence so far is consistent with a Fermi-surface reconstruction caused by the onset of a density-wave order with
a wavevector $Q = (\pi, \pi)$,\cite{lin_theory_2005} 
 which could well be the commensurate N\'eel antiferromagnetic order observed by neutrons at low $x$. \cite{motoyama_spin_2007}
In this case, the pocket at $(\pi,0)$ is electron-like and the pocket at $(\pi/2,\pi/2)$ is hole-like.
One generically expects a Lifshitz transition to occur at a doping well below \xc, where the hole-like pocket disappears,
leaving only the electron-like pocket at $(\pi,0)$ (see Fig.~\ref{PCCO_FSR}).\cite{lin_theory_2005}

The FSR described here will affect all transport properties. 
%\cite{taillefer_fermi_2009}
%
In addition to the dramatic changes in \RH, the resistivity $\rho$ also shows signatures of FSR,
in particular the upturn in $\rho(T)$ seen at low temperature (Fig.~\ref{RvsT}), 
which we attribute to the loss of carrier density. 
%\cite{fournier_insulator-metal_1998}
%
The temperature \Tmin~of the minimum in $\rho(T)$ may then be viewed roughly as the onset of FSR as a function of temperature
(Fig.~\ref{TcTminR30}).
In the resistivity, the onset of FSR at $T=0$ also occurs at \xc~=~0.16, where \Tmin~$\to 0$, 
in agreement with the \RH(0) data, 
but at $T = 20$~K it only occurs at $x = 0.15$ (see Fig.~\ref{TcTminR30}).
In other words, the Fermi surface of PCCO at $x=0.15$ may be considered unreconstructed for $T >$~\Tc.

For our study of superconductivity, what we immediately note is that \Tc~falls with underdoping
as soon as it crosses the \Tmin~line (Fig.~\ref{TcTminR30}).
This strongly suggests that the cause of the \Tc~dome is the FSR -- or, more fundamentally, whatever causes the FSR.
In the following sections, we bring support to this scenario of phase competition in two different ways:
first, by showing that the characteristic field \Hcstar~also falls below $x = 0.15$, forming a dome just like the \Tc~dome;
second, by showing that the superconducting fluctuations on both sides of the dome are not qualitatively different.
These observations remove the need to invoke the emergence of phase fluctuations on the underdoped side.

%------------------------------------------- FIGURE 14 -------------------------------------------------------------------
\begin{figure}
\includegraphics[width=3.5in]{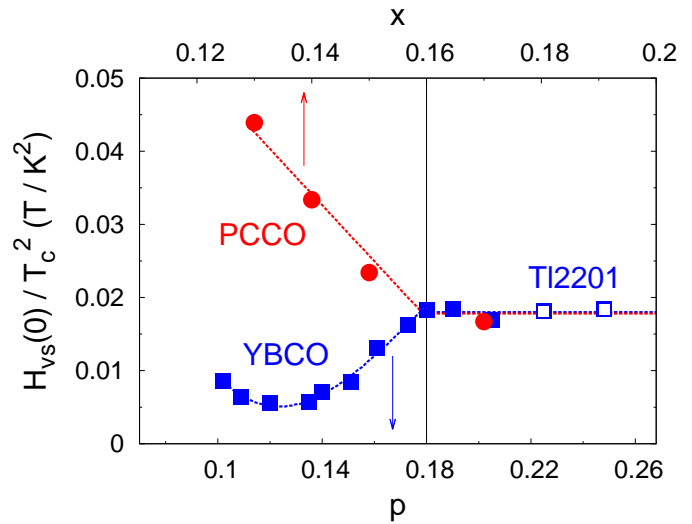}
\caption{\label{Hvs_over_Tc2} 
\Hvs(0)~divided by \Tc$^2$ vs doping, 
for the electron-doped cuprate PCCO (red circles, top horizontal~axis) 
and 
the hole-doped cuprates (bottom horizontal axis; data from ref.~\onlinecite{grissonnanche_direct_2014}) 
YBCO (full blue squares)
and Tl-2201 (open blue squares).
}
\end{figure}
%-------------------------------------------------------------------------------------------------------------------------

\subsection{Characteristic field \Hcstar~and critical field \Hc}
\label{dopingdependenceofHc2}

In Fig.~\ref{x_Hnu_Hstar_Hvs}, we plot \Hcstar~as a function of $x$.
This is the field scale encoded in the superconducting fluctuations just above \Tc.
We see that \Hcstar~tracks \Tc, both showing a dome 
peaking at the same doping, namely $x = 0.15$. 
As already mentioned, this is the doping where FSR onsets for $T \simeq 20$~K (see Fig.~\ref{TcTminR30}).
This shows that what causes \Tc~to fall below $x =0.15$ also causes \Hcstar~to fall,
\ie~the coherence length $\xi$ to increase.
Note that $\xi$ is an average of \vF$ / \Delta$ over the Fermi surface, 
where \vF~is the Fermi velocity and $\Delta$ is the gap magnitude,
so that changes in either \vF~or $\Delta$ (or both) will affect $\xi$, and hence \Hcstar.

Let us examine the evolution of \Hcstar~more closely.
At $T = 20$~K, the Fermi surface is not reconstructed in going from $x = 0.17$ to $x = 0.15$,
so the Fermi-surface average of \vF~should be mostly unchanged.
Therefore the increase in \Hcstar~from 3.0~T to 5.4~T must be due to an increase in $\Delta$.
We can check that by looking at the ratio \Hcstar /\Tc$^2$, which should remain constant if only $\Delta$
changes, since (in a simple model) \Hcstar~$\sim \Delta^2$ and \Tc~$\sim \Delta$.
At $x=0.17$ and $x=0.15$, 
\Hcstar /\Tc$^2 = 17~\pm~2$~mT~/~K$^2$ and  $14~\pm~1$~mT~/~K$^2$, respectively.
We see that within error bars, the rise in both \Hcstar~and \Tc~is driven 
entirely by an increase in $\Delta$.
Moreover, the magnitude of \Hcstar~is consistent with expectation for a dirty $d$-wave superconductor.
Indeed, using \Hcstar~$\simeq \Phi_0 / 2 \pi \xi_0^2$ and expressions for $\xi_0$ given in Appendix~\ref{calculationfromQO}, 
we estimate that \Hcstar~$= 2.3$~T and 5.1~T for $x=0.17$ and 0.15, compared to measured values of
$3.0 \pm 0.3$~T and $5.4 \pm 0.3$~T, respectively.
In summary, we understand the magnitude and doping dependence of 
\Hcstar~in PCCO when its Fermi surface is {\it not} reconstructed.

Let us now see what happens when the Fermi surface {\it is} reconstructed.
The Fermi surface changes from a single large pocket with a large \vF~to
two small pockets with a much smaller \vF. \cite{helm_evolution_2009}
This will boost \Hcstar~and \Hvs(0), so the fact that \Hcstar~nevertheless falls below $x=0.15$ implies that 
$\Delta$ must necessarily decrease.
In Fig.~\ref{Hvs_over_Tc2}, a plot of the ratio \Hvs(0) / \Tc$^2$ vs $x$ reveals that
the enhancement of \Hvs(0)~due to the smaller \vF~gets gradually stronger with underdoping.

Note that when the Fermi surface changes we expect the ratio \Hvs(0) / \Hcstar~to change,
because \Hvs(0) is controlled by those $k$-space regions with the smallest $\xi$,
while \Nsc, and hence \Hcstar, is dominated by those regions with the largest $\xi$ (see Eq.~\ref{serbyn} in Sec.~\ref{gaussiancomparison} and the discussion that follows),
and the relative proportion of these regions will change.
In Fig.~\ref{x_Hnu_Hstar_Hvs}, we see  that while \Hvs(0) = \Hcstar~at $x=0.17$,
the FSR causes \Hvs(0) to become larger than \Hcstar~
and drop at slightly lower doping compared to \Hcstar~due to the details of FSR.
Nevertheless, the main point is that the low value of \Hc~at x = 0.13 is clear evidence that the gap is smaller at that doping (x = 0.13), than it is at x = 0.15.

The emerging picture is the following.
With decreasing $x$, starting at $x \simeq 0.2$, the $d$-wave gap $\Delta_0$ grows, causing \Tc~and \Hc~to grow,
until a critical doping where FSR sets in, whereupon superconductivity is weakened, and both
\Tc~and \Hc~fall.
The FSR is due to the onset of a density-wave state that breaks translational symmetry,
and fundamentally it is this second phase that competes with superconductivity. \cite{taillefer_fermi_2009, taillefer_scattering_2010}
This type of phase competition scenario is observed in several families of unconventional superconductors,
including the quasi-1D organic metals \cite{vuletic_coexistence_2002, doiron-leyraud_correlation_2009}, 
the quasi-2D iron-based superconductors \cite{chubukov_pairing_2012},
and the quasi-3D heavy-fermion metals \cite{mathur_magnetically_1998, gegenwart_quantum_2008},
where in all cases the competing phase is a spin-density wave.
The organizing principle in such a scenario is the QCP where the second phase sets in, invariably
located inside the \Tc~dome. 
We conclude that the fundamental mechanism for a dome of \Tc~vs doping in PCCO is again
phase competition, most likely also with a phase of antiferromagnetic order.\cite{armitage_progress_2010}

\subsection{Comparison to theory of Gaussian fluctuations}
\label{gaussiancomparison}

Until now, we have extracted information from \Nsc~without having recourse to any theory or model or assumption.
We simply obtained \Hstar~directly from the data of \Nsc~vs $H$, and then obtained \Hcstar~directly from the $T$ dependence of \Hstar.
In this section, we compare our data in PCCO to the standard (Aslamazov-Larkin) theory of superconducting fluctuations.  \cite{larkin_theory_2009}

The calculated quantity is the superconducting contribution to the transverse thermo-electric conductivity, \alphaxy, 
while the measured quantity is \Nsc.
The two are related as :
\begin{equation}
\alpha_{\rm xy}^{\rm sc} \simeq \frac{N_{\rm sc}}{\rho} ~~~,
\label{USHsimple}
\end{equation}
assuming $|S\tan{\theta_H}|\ll N$ (see Appendix~\ref{StoN}).
In 2002,
Ussishkin, Sondhi, and Huse calculated the thermo-electric response of a quasi-2D type-II superconductor in the Gaussian approximation,
in the limits of $H \to 0$ and $T \to$~\Tc.\cite{ussishkin_gaussian_2002}
In 2009, calculations of \alphaxy~in a dirty 2D type-II superconductor were extended to arbitrary $T$ and arbitrary $H$
by two groups independently,\cite{michaeli_fluctuations_2009, serbyn_giant_2009} 
who arrived at similar results.
We now compare the magnitude and field dependence of \Nsc~measured in PCCO with the latest predictions of Gaussian theory.

\subsubsection{Magnitude}

In the limit of $H \to 0$ and $T \to$~\Tc, \alphaxy~above \Tc~is given, in two dimensions, by: \cite{serbyn_giant_2009}
\begin{equation}
\alpha_{\rm xy}^{\rm sc} = \frac{2}{3} \frac{k_{\rm B}e}{h}  \frac{H}{\tilde{H}_{\rm c2}(0)}   \frac{1}{\epsilon}~~~,
\label{serbyn}
\end{equation}
where $k_{\rm B} e/h = 3.33$~nA/K is the quantum of thermoelectric conductance, \cite{pourret_observation_2006, pourret_nernst_2009}
$\tilde{H}_{\rm c2}(0) = \Phi_0 / 2 \pi \xi^2(0)$, and $\epsilon = (T-T_{\rm c})/T_{\rm c}$.
The magnitude of \alphaxy~is seen to depend on one quantity only, the Ginzburg-Landau coherence length $\xi(0)$,
so that \alphaxy~$\sim \xi^2(0)$ as mentioned in Sec.~\ref{dopingdependenceofHc2}.
To make contact with experiment, we use the relation \Hc~$\simeq 0.59~\tilde{H}_{\rm c2}(0)$,  \cite{larkin_theory_2009}
 and the empirical facts that \Hc(0)~=~\Hvs(0) and \Hvs(0)~=~\Hcstar~(for a single large circular Fermi surface).
This yields a theoretical expression where the only input parameter is \Hcstar:
\begin{equation}
{Theory :}~~~~~~\frac{\alpha_{\rm xy}^{\rm sc}}{H} \simeq~0.4~\frac{k_{\rm B}e}{h} \frac{1}{H_{\rm c2}^\star} \frac{1}{\epsilon}~~.
\label{serbyn_modified}
\end{equation}
The measured value of \alphaxy~is determined using:
\begin{equation}
{Experiment :}~~~~~~\frac{\alpha_{\rm xy}^{\rm sc}}{H} \simeq \frac{\nu_0^{\rm sc}}{\rho_\square}~~~~~~~~~~,
\label{alpha_experiment}
\end{equation}
where $\nu_0^{\rm sc} = \nu_0 - \nu_0^{\rm qp}$ is the superconducting Nernst coefficient in the $H \to 0$ limit (Fig.~\ref{nu0}), 
and $\rho_\square = \rho/s$, in terms of the zero-field electrical resistivity $\rho$ (Fig.~\ref{RvsT}) and the interlayer separation
$s = 6.1$~\AA.

%------------------------------------------- FIGURE 15 --------------------------------------------------------------------
\begin{figure}
\includegraphics[width=3.5in]{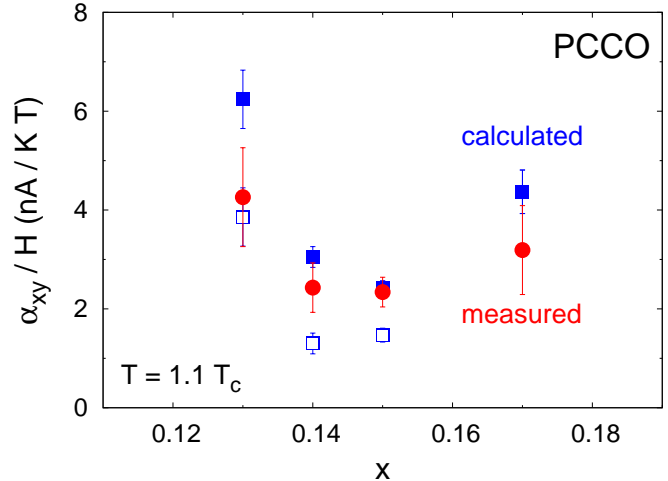}
\caption{\label{alpha_xy}
Comparison of measured (full red circles; Eq.~\ref{alpha_experiment}) and calculated 
(full blue squares; Eq.~\ref{serbyn_modified}) 
values of \alphaxy~$/ H$ vs $x$.
Open blue squares are the values calculated using \Hvs(0) instead of \Hcstar~in Eq.~\ref{serbyn_modified}.
The agreement between experiment and theory, with no adjustable parameter, is remarkable.
}
\end{figure}
%-------------------------------------------------------------------------------------------------------------------------

In Fig.~\ref{alpha_xy},
we plot the theoretical and experimental values of \alphaxy/$H$ at $\epsilon=0.1$,
using the values of \Hcstar~given in Table~\ref{scaling}
and the measured values of $\nu_0^{\rm sc}$ 
and $\rho$ given in Table~\ref{alphatable}, respectively.
The agreement between theory and experiment is remarkable.
Although a number of factors not considered here (\eg~$s$-wave vs $d$-wave) could alter this quantitative agreement somewhat,
it is nevertheless evident that Gaussian theory can reliably explain not only the magnitude of \Nsc~in PCCO, but also its detailed doping dependence.
In particular, it shows that there is no qualitative difference between the superconducting fluctuations
of the underdoped regime relative to the overdoped regime. 
The fluctuations are Gaussian everywhere, meaning that the superconducting order parameter fluctuates
in both amplitude and phase in the same way across the phase diagram.
This therefore rules out the long-held notion that phase fluctuations play a special role in underdoped cuprates. \cite{Emery_1995}

%------------------------------------------- FIGURE 16 ---------------------------------------------------------------------
\begin{figure}
\includegraphics[width=3.5in]{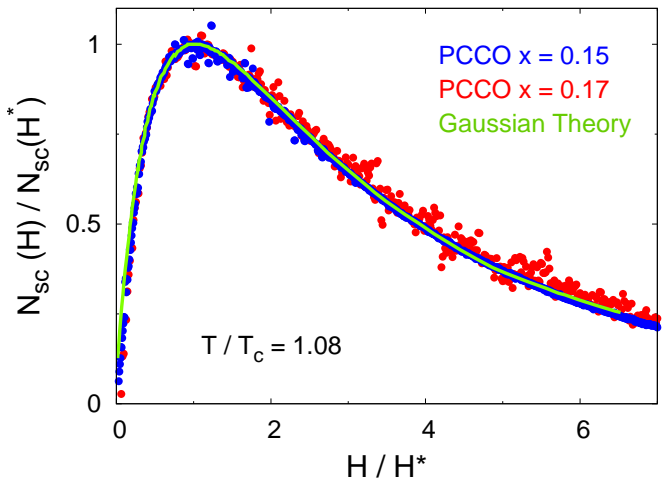}
\caption{\label{gaussian} 
Superconducting Nernst signal of PCCO in the absence of FSR,
plotted as a function of magnetic field for the overdoped ($x=0.17$; red) and optimally doped ($x=0.15$; blue) samples,
measured at $T = 1.08$~\Tc~in both cases.
The data are compared to the theoretical calculation of \alphaxy~(green curve, from ref.~\onlinecite{michaeli_fluctuations_2009}).
For both experimental data and calculated curve, 
the $x$ axis is normalized by the peak field \Hstar~and the $y$ axis is normalized by the magnitude of the superconducting Nernst signal at \Hstar. 
The agreement between theory and experiment is excellent.}
\end{figure}
%-------------------------------------------------------------------------------------------------------------------------

\subsubsection{Field dependence}

One may ask whether the fluctuations at high field might be different from those close to $H = 0$~that 
were considered in the previous section.
In the conventional superconductor \NbSi, 
direct comparison\cite{michaeli_fluctuations_2009} of the calculated \alphaxy~and the measured \Nsc~(refs.~\onlinecite{pourret_observation_2006, pourret_length_2007}) showed detailed quantitative agreement,
validating the theory of Gaussian fluctuations at arbitrary $H$.

In Fig.~\ref{gaussian}, we reproduce the calculated curve of \alphaxy~vs $H$ at $T = 1.08$~\Tc.\cite{michaeli_fluctuations_2009}
It shows the characteristic rise and fall, with a peak at some field \Hstar. 
The curve is normalized at \Hstar~for both axes, allowing us to  compare with our data for \Nsc~vs $H$, normalized in the same way.
In Fig.~\ref{gaussian}, we perform this comparison for the two dopings at which the Fermi surface is not reconstructed,
namely $x = 0.15$ and $x = 0.17$.
The data are seen to be in perfect agreement with the theoretical curve, for both dopings.
This shows that the theory of Gaussian fluctuations continues to be valid in PCCO well beyond the
limit of small fields.

We conclude that all aspects of our data in PCCO agree with the theory of Gaussian fluctuations.

Before moving on to the next section, let us comment on the behavior of \Hstar~close to \Tc.
At all dopings, we see that \Hstar~deviates from its $\ln(T/T_{\rm c})$ dependence as $T \to$~\Tc,
in such a way that \Hstar~saturates to a non-zero value at $T =$~\Tc, or $\epsilon = 0$ (Fig.~\ref{T_Hstar}).
This is a reflection of the fact that the initial rise in \Nsc~vs $H$ never becomes infinitely rapid, 
even at $\epsilon = 0$~(Fig.~\ref{NvsH_alldopings}, right panels).
As can be seen in Fig.~\ref{nu0}, the initial slope $\nu_0$ does not diverge as $T \to$~\Tc, for any doping.
On the contrary, $\nu_0$ saturates to a constant value below $\epsilon \simeq 0.1$.
This saturation is entirely expected on theoretical grounds, since $\nu \sim$~\alphaxy$ / \sigma$ is the ratio
of two coefficients that both diverge in the same way as $\epsilon \to 0$. 
Indeed, just as \alphaxy~$\propto 1 / \epsilon$ (Eq.~\ref{serbyn}), so is $\sigma \propto 1 / \epsilon$.  \cite{larkin_theory_2009}

\subsection{Comparison with hole-doped cuprates}
\label{comparisonwithholedoped}

We have shown that in the electron-doped cuprate PCCO the superconducting fluctuations are Gaussian
throughout the doping phase diagram, ruling out phase fluctuations as a mechanism for the \Tc~dome,
and superconductivity weakens as soon as Fermi-surface reconstruction sets in, below a critical doping \xc.
The origin of the \Tc~dome is therefore an underlying growth in the gap $\Delta_0$ with decreasing $x$,
curtailed by the onset of a competing phase.
This is why the dome is centered around \xc.
In this section, we investigate to what extent a similar scenario applies to hole-doped cuprates.

\subsubsection{FSR and the origin of the \Tc~dome}

The first thing to note is that hole-doped cuprates also undergo a FSR below some critical doping \pc.\cite{taillefer_fermi_2009, taillefer_scattering_2010}
This was revealed unambiguously by the discovery of low-frequency quantum oscillations in YBCO,\cite{doiron-leyraud_quantum_2007}
shown to come from a small electron-like pocket in the Fermi surface of underdoped YBCO,
because of the large negative Hall coefficient \RH~at low temperature.\cite{leboeuf_electron_2007}
In the normal state, once superconductivity has been removed by application of a large magnetic field,
the electron pocket is seen to persist as a function of doping up to at least $p=0.15$.\cite{leboeuf_lifshitz_2011}
Given that hole-doped cuprates above $p \simeq 0.25$
are known to have a single large hole-like Fermi surface \cite{mackenzie_normal-state_1996, plate_fermi_2005, vignolle_quantum_2008} 
(very similar 
to that of electron-doped cuprates at high $x$), 
the FSR in the normal state at $T=0$ must take place at a critical doping \pc~such that $0.15 <$~\pc~$<0.25$.

%------------------------------------------- FIGURE 17 ----------------------------------------------------------------------
%\begin{figure}
%\includegraphics[width=3.5in]{FIGURES/YBCO_Tx}
%\caption{\label{YBCO_Tx} 
%
%Temperature-doping phase diagram of YBCO, showing the superconducting dome delineated by the zero-field
%critical temperature \Tc~(black circles).
%
%Also shown is \Tx~(red squares), the temperature at which the $a$-axis resistivity $\rho_{\rm a}(T)$ has an
%inflection point. \cite{ando_electronic_2004}
%
%The red line is a linear fit to the \Tx~data, extrapolated to $T=0$ (open square).
%
%The corresponding doping, \pc~$\simeq 0.19$, is the quantum critical point below which the 
%Fermi-surface reconstruction onsets,
%consistent with the critical doping where the normal-state Hall coefficient \RH~at $T \to 0$ changes sign,
%from positive above $p \simeq 0.25$ to negative below 
%$p = 0.15$.\cite{leboeuf_lifshitz_2011}
%}
%\end{figure}
%-------------------------------------------------------------------------------------------------------------------------

In YBCO, the onset of this FSR upon cooling is rather gradual, as it is in PCCO, and it may be said to occur
at the temperature \Tmax~below which \RH$(T)$ starts to fall towards negative values.\cite{leboeuf_lifshitz_2011}
The fact that the \Tmax~line and the \Tc~line cross where the latter peaks is strong evidence that the
the drop of \Tc~on the underdoped side is linked to the FSR.\cite{leboeuf_lifshitz_2011, grissonnanche_direct_2014}

%In YBCO, the FSR does not cause an upturn in $\rho(T)$, although it does in other hole-doped cuprates,
%like \NdLSCO~(Nd-LSCO).\cite{ichikawa_local_2000, daou_linear_2009}
%
%Whether FSR produces an upturn or a downturn depends on the relative strength of inelastic and 
%elastic scattering.
%
%In the iron-based superconductor BaFe$_2$As$_2$, for example, the FSR caused by the onset of antiferromagnetic order 
%at \TN~yields
%an upturn in $\rho(T)$ if the material is doped with Co and a downturn if doped with K.\cite{doiron-leyraud_quantum_2012}
%
%In case of a downturn, we use the inflection point in $\rho(T)$, at a temperature \Tx, as the signature of the FSR.
%In K-doped BaFe$_2$As$_2$, \Tx~$\simeq$~\TN.
%
%Ando and co-workers have performed extensive measurements of the $a$-axis resistivity $\rho_a(T)$ in YBCO; \cite{ando_electronic_2004}
%we reproduce their data for \Tx~vs $p$ in Fig.~\ref{YBCO_Tx}.

The temperature-doping phase diagram of YBCO is therefore similar to that of PCCO (Fig.~\ref{TcTminR30}),
in the sense that the onset of FSR
%, at \Tx~or \Tmin, 
extrapolates to a critical point at $T=0$ which lies just above optimal
doping (where \Tc~peaks).
%In YBCO, \Tx~extrapolates to 
%$p \simeq 0.19$, in agreement with the interval imposed by Fermi-surface measurements.
%
So the origin of the \Tc~dome in hole-doped cuprates appears to be fundamentally the same as in electron-doped cuprates,
namely phase competition and FSR below a quantum critical point located inside the dome.
Note, however, that the competing order itself may be different.

\subsubsection{Critical field \Hc~and critical doping \pc}
\label{YBCO_pc}

\begin{figure}
\includegraphics[width=3.5in]{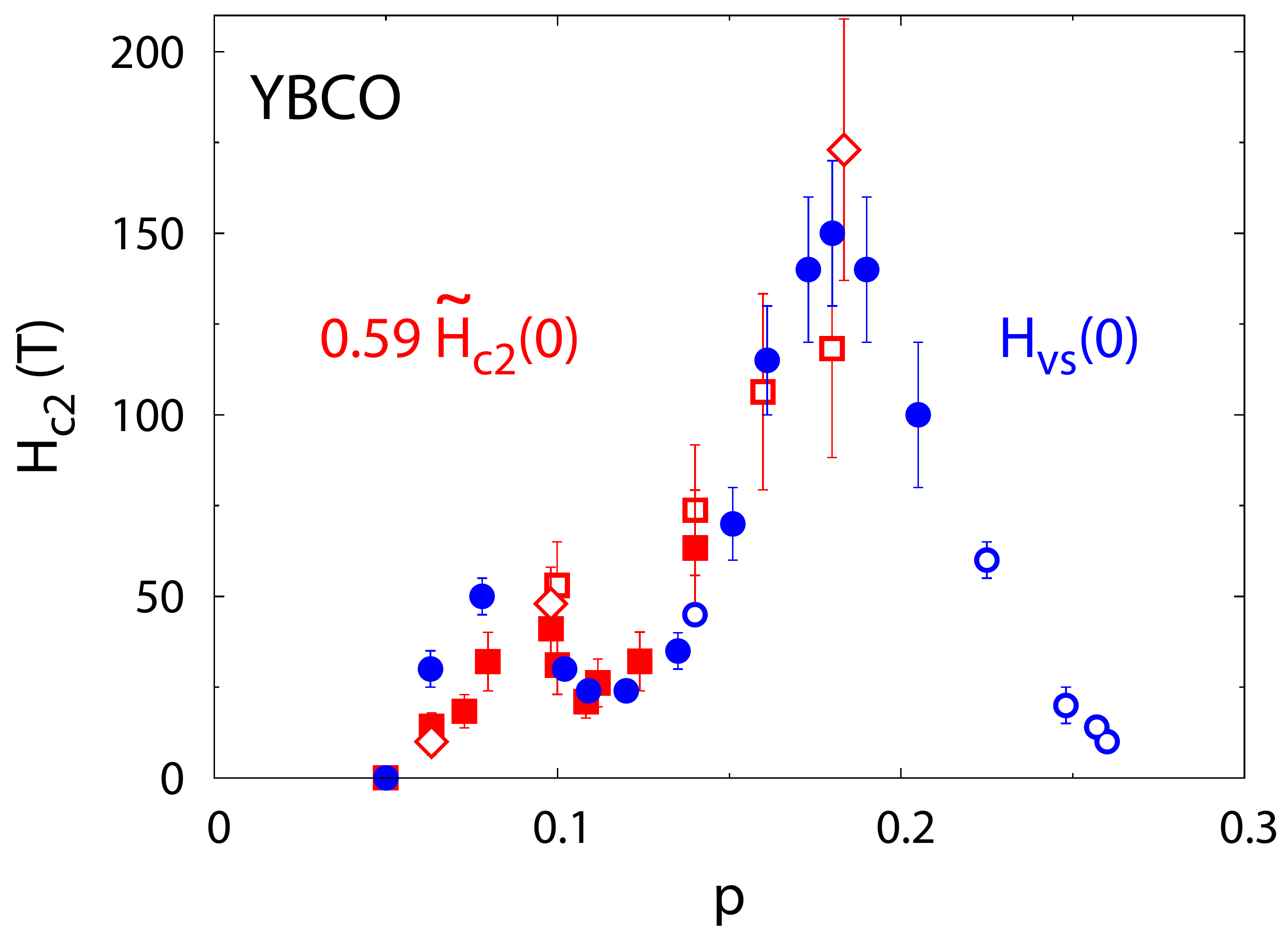}% Here is how to import EPS art
\caption{\label{YBCO} 
Upper critical field \Hc~in the hole-doped cuprate YBCO, obtained in two different ways.
First, directly from resistive measurements of \Hvs($T$)~at low temperature in high magnetic fields,
extrapolated to $T=0$, giving \Hc~=~\Hvs(0) (blue symbols).\cite{grissonnanche_direct_2014}
The open  circles are for YBa$_2$Cu$_4$O$_8$ ($p=0.14$)
 and Tl-2201 ($p >Ê0.21$).\cite{grissonnanche_direct_2014}
Second, we plot $0.59~\tilde{H}_{\rm c2}(0)$ (red symbols), with $\tilde{H}_{\rm c2}(0) = \Phi_0 / 2 \pi \xi^2(0)$,
where the coherence length $\xi(0)$ is obtained from Gaussian Aslamazov-Larkin theory 
applied above \Tc~to either the conductivity (full squares from ref.~\onlinecite{ando_magnetoresistance_2002},
open squares from ref.~\onlinecite{rullier-albenque_high-field_2011})
or the magnetization (open diamonds, from ref.~\onlinecite{kokanovic_diamagnetism_2013}).
The agreement between the two ways of determining \Hc~is remarkable.
}
\end{figure}

To investigate the comparison further, we examine the doping dependence of critical fields in hole-doped cuprates.
The upper critical field \Hc~was recently determined by thermal conductivity
measurements in YBCO, 
YBa$_2$Cu$_4$O$_8$ and 
\Tl~(Tl-2201). \cite{grissonnanche_direct_2014}
The data revealed that \Hc~=~\Hvs(0), as confirmed 
here in PCCO at $x=0.15$.
Using high-field measurements of \Hvs($T$), the complete doping dependence of \Hc~was reported;
the data are reproduced in Fig.~\ref{YBCO}.
We see that \Hc~vs $p$ 
exhibits two peaks, pointing to two underlying quantum critical points,
possibly associated with the onset of two distinct competing phases.\cite{grissonnanche_direct_2014}
Here we focus on the higher peak, at \pc~=~0.18.
Starting from high doping, we see that \Hc~rises from zero at $p \simeq 0.27$ up to \Hc~$= 150 \pm 20$~T at \pc.
In this overdoped regime, the ratio \Hc/\Tc$^2$ is roughly constant (Fig.~\ref{Hvs_over_Tc2}), showing that 
the growth in the gap magnitude $\Delta_0$~controls the rise of \Tc~and \Hc, 
as we found in overdoped PCCO.

%------------------------------------------- FIGURE 18 ---------------------------------------------------------------------
%-------------------------------------------------------------------------------------------------------------------------

Moreover, the value of the ratio is 
roughly the same in both hole-doped and electron-doped materials, 
namely 
\Hc/\Tc$^2 \simeq 18$~mT/K$^2$~(Fig.~\ref{Hvs_over_Tc2}). 
This may be somewhat coincidental, since \Hc~depends on the Fermi velocity \vF~and the mean free path $l$~(see Appendix),
and these parameters may not be identical in YBCO or Tl-2201 and in PCCO, but it nevertheless explains
why \Hc~in PCCO is so much smaller than in YBCO.
Indeed, a factor 5 smaller \Tc, from \Tc~$\simeq 100$~K to \Tc~$\simeq 20$~K, will yield a factor 25 smaller \Hc, 
from \Hc~$\simeq 150$~T to \Hc~$\simeq 6$~T,
as roughly observed.

Below \pc~=~0.18, \Hc~in YBCO drops by a factor 6, down to \Hc~$= 24 \pm 2$~T at $p=0.11$.\cite{grissonnanche_direct_2014}
The condensation energy $\delta E$ drops by a factor 20, and the ratio $\delta E / T_{\rm c}^2$ by a factor 8.\cite{grissonnanche_direct_2014}
This dramatic suppression of superconductivity is attributed to phase competition,
involving the onset of charge order. \cite{grissonnanche_direct_2014, chang_direct_2012, ghiringhelli_long-range_2012}
The value of \pc~is consistent with the onset of FSR at $T=0$, as estimated from \RH. \cite{leboeuf_lifshitz_2011}
%~or from \Tx~(Fig.~\ref{YBCO_Tx}).

As seen in Fig.~\ref{Hvs_over_Tc2}, the FSR in YBCO causes a {\it drop} in the ratio \Hc/\Tc$^2$,
at least initially, whereas the FSR in PCCO causes an increase.
The difference is likely to come at least in part from the different effect of FSR on the Fermi velocity \vF.
We mentioned that in PCCO \vF~undergoes a large change, by an order of magnitude.
In YBCO, however, quantum oscillation measurements
give only a factor 2 drop in \vF~upon FSR, if we compare overdoped Tl-2201 ($p \simeq 0.25$; ref.~\onlinecite{bangura_fermi_2010}) 
and underdoped YBCO ($p = 0.11$; ref.~\onlinecite{jaudet_haasvan_2008}).
The difference could also come from a different topology of the reconstructed Fermi surface.
While the electron pocket in the Fermi surface of PCCO is located at $(\pi, 0)$, where the $d$-wave gap 
is maximal, the electron pocket in the Fermi surface of YBCO is quite possibly located at
$(\pi/2, \pi/2)$,
where the gap goes to zero. \cite{allais_connecting_2014}

%------------------------------------------- FIGURE 19 --------------------------------------------------------------------
\begin{figure}
\includegraphics[width=3.5in]{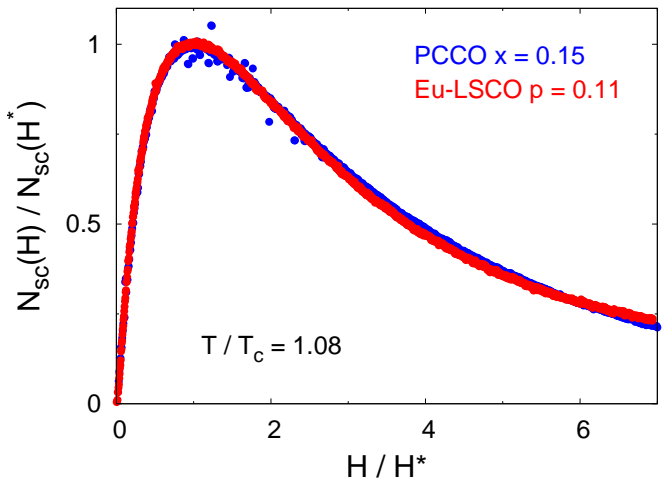}
\caption{\label{holedoped_scaled} 
Superconducting Nernst signal of electron-doped PCCO (blue)
and hole-doped Eu-LSCO (red), measured at $T = 1.08$~\Tc~in both cases,
for dopings as indicated.
The $x$ axis is normalized by the peak field \Hstar~and the $y$ axis is normalized by the magnitude of the superconducting Nernst signal at \Hstar. 
The agreement between the two is excellent,
showing that the nature of superconducting fluctuations is basically the same
whether cuprates are electron-doped or hole-doped,
and both are well described by Gaussian theory (see Fig.~\ref{gaussian}).
}
\end{figure}
%-------------------------------------------------------------------------------------------------------------------------

\subsubsection{Superconducting fluctuations}

Superconducting fluctuations in hole-doped cuprates have been studied extensively,
mostly via measurements of the electrical resistivity, the magnetization and the Nernst effect.
An exhaustive study of paraconductivity in YBCO by Ando and co-workers showed that Gaussian theory
(Aslamazov-Larkin) accounts well for the effect of superconducting fluctuations above \Tc. \cite{ando_magnetoresistance_2002}
From their analysis, justified in light of other works,\cite{ramallo_-plane_1996, curras_plane_2003}
they extract a coherence length $\xi$ as a function of doping,
and use it to estimate the critical field 
$\tilde{H}_{\rm c2}(0) = \Phi_0 / 2 \pi \xi^2$.
A later study by Rullier-Albenque and co-workers, based on a similar analysis,
yielded $\tilde{H}_{\rm c2}(0)$~values in good agreement with the earlier work, at least for $p < 0.15$. \cite{rullier-albenque_high-field_2011}
For $p > 0.15$, the use of higher magnetic fields in the more recent study may have improved the estimate of the 
underlying normal-state magneto-resistance. \cite{rullier-albenque_total_2007}
In Fig.~\ref{YBCO}, we plot 
\Hc~$\simeq 0.59~\tilde{H}_{\rm c2}(0)$ obtained from the data of both groups
on the 
$H$-$p$ diagram of YBCO.
We also plot \Hc~obtained from recent magnetization measurements analyzed using  
Gaussian theory to extract the coherence length. \cite{kokanovic_diamagnetism_2013}.
The agreement between \Hc~obtained from direct measurements of \Hvs~at low temperature and high fields
and \Hc~encoded in the superconducting fluctuations above \Tc~is remarkable.
In particular, it exhibits the same two-peak structure and the same six-fold drop between $p=0.18$ and $p=0.11$.

\begin{figure}
\includegraphics[width=3.5in]{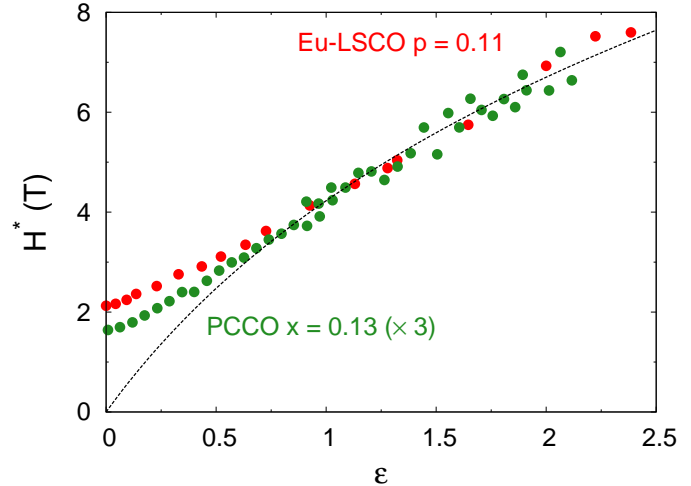}
\caption{\label{Hstar_PCCOvsEuLSCO} 
Comparison of \Hstar~vs $\epsilon$~for electron-doped PCCO (green circles)
and hole-doped Eu-LSCO (red circles),
at dopings as indicated.
The PCCO data are multiplied by a factor 3.
The black dotted line is a fit of the data to the function \Hstar = \Hcstar~$\ln (T/T_{\rm c})$~(Eq.~\ref{Hstar}).
}
\end{figure}

The Nernst response of hole-doped cuprates was recently revisited 
and shown to be in excellent agreement with the Gaussian theory. \cite{chang_decrease_2012}
Figs.~\ref{holedoped_scaled} and  \ref{Hstar_PCCOvsEuLSCO} show how the behavior of \Nsc~in 
the hole-doped cuprate Eu-LSCO is qualitatively identical to that of electron-doped PCCO,
as a function of field and temperature, respectively.

We conclude that superconducting fluctuations in cuprates are well described by
Gaussian theory, whether in electron-doped or hole-doped materials,
whether in the overdoped or underdoped regimes.
This may no longer be true at very low doping, \ie~$p < 0.08$ and $x < 0.13$, when close to the Mott insulator,
but otherwise it appears that using the standard theory is a reliable way to extract fundamental information
about superconductivity in the cuprates. 
Previous analyses of paraconductivity and diamagnetism in hole-doped cuprates like YBCO have also found Gaussian theory to be a good description of the superconducting fluctuations. \cite{curras_plane_2003, ramallo_comment_2012, rey_comment_2013}

%------------------------------------------- FIGURE 20 -------------------------------------------------------------------
%-------------------------------------------------------------------------------------------------------------------------

%%%%%%%%%%%%%%%%%%%%%%%%%%% %%%%%%%%%%%%%%%%%%%%%%%%%%%%%%%%%%
%%%%%%%%%%%%%%%%%%%%    SUMMARY   %%%%%%% %%%%%%%%%%%%%%%%%%%%%%%%%%%
 %%%%%%%%%%%%%%%%%%%%%%%%%%% %%%%%%%%%%%%%%%%%%%%%%%%%%%%%%%%%%

\section{\label{conclusion}Summary}

Our measurements of the Nernst effect in the electron-doped cuprate superconductor PCCO
elucidate the nature of superconducting fluctuations above the critical temperature \Tc.
We find that the superconducting Nernst signal \Nsc~is in qualitative and quantitative agreement with 
the theory of Gaussian fluctuations in dirty 2D superconductors,\cite{ussishkin_gaussian_2002, serbyn_giant_2009, michaeli_fluctuations_2009}
at all dopings.
Indeed, \Nsc$(T, H)$ in PCCO behaves as it does in the conventional superconductor \NbSi. \cite{pourret_observation_2006, pourret_length_2007}
This implies that there is nothing unusual about the fluctuations in PCCO,
even in the underdoped regime.
There is no evidence of any vortex-like excitations, or pre-formed pairs, above \Tc.

The characteristic magnetic field scale \Hcstar~extracted directly from the data independent of any theoretical assumptions
displays a dome-like dependence on doping, showing the same $x$ dependence as \Tc,
both peaking at $x = 0.15$.
This shows that the pairing strength drops below optimal doping.
This weakening of superconductivity occurs at the critical doping where the Fermi surface
undergoes a reconstruction.\cite{dagan_evidence_2004} 
The scenario is therefore one of competition with another ordered phase 
that causes both the Fermi-surface reconstruction and the suppression of \Tc~and \Hcstar. {\cite{taillefer_fermi_2009, taillefer_scattering_2010, 
doiron-leyraud_quantum_2012,
grissonnanche_direct_2014,
alloul_superconducting_2010}
Most likely, the competing phase in PCCO is antiferromagnetic order.\cite{lin_theory_2005} 

The emerging picture for PCCO is therefore the same as in quasi-1D organic superconductors,
where an antiferromagnetic quantum critical point is clearly the organizing principle. \cite{doiron-leyraud_correlation_2009, taillefer_scattering_2010}
The magnetic fluctuations cause $d$-wave pairing while the magnetic order competes
with superconductivity. The first effect increases \Tc, the second decreases \Tc, and the
two together produce the \Tc~dome that straddles the critical point.
By analogy, we infer that the same two mechanisms are at play in PCCO,
supporting the case for magnetically-mediated pairing in cuprates. \cite{jin_link_2011, taillefer_scattering_2010}
This is also the likely scenario in iron-based superconductors.\cite{chubukov_pairing_2012}

A comparison of our Nernst data on PCCO with corresponding data reported for the hole-doped 
cuprate Eu-LSCO (ref.~\onlinecite{chang_decrease_2012}) reveals a strong similarity. 
We conclude that superconducting fluctuations in hole-doped cuprates are not significantly
different. 
This validates the use of Gaussian fluctuation theory in previous analyses of paraconductivity data in underdoped YBCO,
which yielded an estimate of \Hc~that decreases rapidly with underdoping,\cite{ando_magnetoresistance_2002, rullier-albenque_high-field_2011}
in good agreement with direct measurements of \Hc~in YBCO.\cite{grissonnanche_direct_2014}
These studies establish that the dominant mechanism for the \Tc~dome in hole-doped cuprates
is also phase competition.
Interestingly, in this case the competition does not seem to come from magnetic order,
but appears to involve charge order. \cite{ghiringhelli_long-range_2012, chang_direct_2012}

We conclude that fluctuations in the phase of the superconducting order parameter, long invoked
as the mechanism responsible for the \Tc~dome of cuprates, 
do not in fact play a prominent role in the origin of the \Tc~dome of either electron-doped
or hole-doped cuprates.

\begin{acknowledgments}

We thank 
H.~Alloul,
P.~Armitage,
Y.~Ando,
K.~Behnia,
A.~Carrington,
J.~Chang,
G.~Deutscher,
N.~Doiron-Leyraud,
V.~Galitski,
R.~L.~Greene,
S.~A. Kivelson,
P.~A.~Lee,
K.~Michaeli,
M.~R.~Norman,
M.~Ramallo,
M.~Serbyn,
J.~Sonier,
A.~Varlamov,
and
F.~Vidal
for stimulating discussions, 
and
J.~Corbin, S.~Fortier, F.~Francoeur and
A.~Mizrahi
for assistance with the experiments.
This work was supported by the Canadian Institute for Advanced Research 
and a Canada Research Chair, and it was funded by NSERC, FRQNT and CFI.

\end{acknowledgments}

%%%%%%%%%%%%%%%%%%%%%%%%%%% %%%%%%%%%%%%%%%%%%%%%%%%%%%%%%%%%%
%%%%%%%%%%%%%%%%%%%%    APPENDIX   %%%%%%%%%%%%%%%%%%%%%%%%%%%%%%%%%%
 %%%%%%%%%%%%%%%%%%%%%%%%%%% %%%%%%%%%%%%%%%%%%%%%%%%%%%%%%%%%%

\appendix

\section{\label{calculationfromQO}
Estimate of \Hc~in PCCO}

Quantum oscillations in the $c$-axis resistivity of overdoped NCCO at $x=0.17$ have a frequency $F=10960$~T and 
a cyclotron mass $m^\star = 2.3~m_0$.\cite{helm_evolution_2009}
Using the Onsager relation: 
\begin{equation}
\label{onsager}
F = \frac{\Phi_0}{2\pi^2} A_k
\end{equation}
with $\Phi_0 = 2.07 \times 10^{-15}$~Wb, 
we extract the corresponding values for the Fermi wavevector of the large Fermi surface $k_{\rm F} = 0.58$~\AA$^{-1}$ 
and the Fermi velocity $v_{\rm F} = \hbar k_{\rm F} / m^\star = 2.9 \times 10^5$~m/s.   
Using the following relation for the coherence length of a clean superconductor:
\begin{equation}
\label{coherence}
\xi_0^{\rm clean}=\frac{\hbar v_F}{a\Delta_0}
\end{equation}
with $a = 1.5$ and $\Delta_0 = 2.14~k_{\rm B}$\Tc~for a $d$-wave state, 
we extract the clean limit coherence length $\xi_0^{\rm clean}=51$~nm. 
Disorder affects the clean limit coherence length if the mean free path $\ell$ is comparable to $\xi_0^{\rm clean}$.
We estimate the mean free path via the relation: 
\begin{equation}
\label{meanfree}
\ell = \frac{hs}{e^2\rho_0k_F}
\end{equation}
where $\rho_0$ is the residual resistivity.
Using the inter-layer distance $s = 6.1$~\AA~and $\rho_0 = 17.5~\mu \Omega$~cm (Fig. \ref{RvsT}),
we get $\ell=15.5$~nm.  
Using Pippard's relation for the coherence length of dirty superconductors:
\begin{equation}
\label{pippard}
\frac{1}{\xi_0^{\rm dirty}}=\frac{1}{\xi_0^{\rm clean}}+\frac{1}{\ell}~~~~,
\end{equation}
we arrive at a coherence length of $\xi_0^{\rm dirty}=11.9$~nm in the dirty limit.
We use this value to calculate the upper critical field from the expression:
\begin{equation}
\label{hcxi}
H_{\rm c2} = \frac{\Phi_0}{2 \pi \xi^2_0} = 2.3~\rm{T}~~~~.
\end{equation}
The result is close to our measured value $H_{\rm{vs}}(T\to0)=3.0\pm0.2$~T, as discussed in Sec.~\ref{dopingdependenceofHc2} and presented in Table~\ref{scaling}.

%%%%%%%%%%%%%%%%%%%%%     TABLE III     %%%%%%%%%%%%%%%%%%%%%
\begin{table}[b]%The best place to locate the table environment is directly after its first reference in text
\caption{\label{stonratio} Comparing the magnitude of $|S\tan{\theta_{\rm H}}|$ relative to $N$ in our PCCO samples. }
\begin{tabular}{c c c c c}
\hline
\textrm{$x$}&
\textrm{$T_c$}&
\textrm{$|S\tan(\theta_H)|$}&
\textrm{$N$}&
\textrm{$\frac{|S\tan(\theta_H)|}{N}$}\\
       &
   (K) &
(nV/K) &
(nV/K) &  
       \\ 
\hline
\hline
0.13 & 8.8 & 36 & 263 & 0.14\\
0.14 & 17.4 & 11 & 1394 & 0.01\\
0.15 & 19.5 & 1 & 1359 & 0.001\\
%0.16 & 14.3 & 3.8 & 3.9 & 0.02 & 0.2 & 90.2\\
0.17 & 13.4 & 17 & 425 & 0.04\\
\hline
\end{tabular}
\end{table}
%%%%%%%%%%%%%%%%%%%%%%%%%%      TABLE          %%%%%%%%%%%%%%%%%%%%%%%%%

\section{\label{StoN}Magnitude of $S\tan(\theta_{\rm H})$}

Eq. \ref{USHsimple} is valid when $|S\tan{\theta_{\rm H}}|\ll N$.
To verify that this condition is indeed satisfied in our PCCO samples, we have measured the Seebeck effect in our thin films and used the values of $R_H$ from Ref.~[\onlinecite{charpentier_antiferromagnetic_2010}] to calculate $\tan(\theta_H)$ using $\tan(\theta_{\rm H})=R_{\textrm H}B/\rho$.
Table~\ref{stonratio} lists the value of $|S\tan{\theta_{\rm H}}|$, $N$, and the ratio of the two quantities at $T=4$~K and $H=15$~T for our four samples.
The condition $|S\tan{\theta_{\rm H}}|\ll N$ is seen to hold at all dopings.

% The \nocite command causes all entries in a bibliography to be printed out
% whether or not they are actually referenced in the text. This is appropriate
% for the sample file to show the different styles of references, but authors
% most likely will not want to use it.
%\nocite{*}
%%%%%%%%%%%%%%%%%%%%%%%%%%%% BIBLIOGRAPHY

\bibliography{PCCO_17july2014}

% ADDITIONAL REFERENCES

% dagan_2004     %% PRL 2004 on QCP in PCCO

% kapitulnik_1985   %% Physica paper with Deutscher

\end{document}